\documentclass[twocolumn,final,natbib]{svjour3}
\newif\ifprep \preptrue\newif\iffinal\finaltrue    
\usepackage{graphicx}
\usepackage{natbib}
\journalname{Journal of Geodesy}

\newcommand{\vex}{\vspace{1ex}}
\newcommand{\hps}{\hphantom{)}}
\newcommand{\hpl}{\hphantom{+}}
\newcommand{\ntab}[2]{ \multicolumn{1}{#1}{#2} }
\newcommand{\nntab}[2]{ \multicolumn{2}{#1}{#2} }
\newcommand{\nnntab}[2]{ \multicolumn{3}{#1}{#2} }

\newcommand{\dss}{\displaystyle}

\newcommand{\const}{\rm const}
\newcommand{\hp}{\hphantom{-}}
\newcommand{\mat}[1]{\widehat{\mathstrut\cal #1}}

\newcommand{\tram}[1]{\mat{#1}\hspace{-0.05em}\raisebox{1.1ex}{\bf\tiny $\top$}\hspace{-0.2em}}
\newcommand{\trav}[1]{\vec{#1}\hspace{-0.05em}\raisebox{1.1ex}{\bf\tiny $\top$}\hspace{-0.2em}}
\newcommand{\Cov}{ \mathop{ \rm Cov }\nolimits }
\newcommand{\nc}[1]{ \multicolumn{1}{c}{#1} }
\newcommand{\nl}[1]{ \multicolumn{1}{l}{\hspace{-1em}#1} }
\newcommand{\Az}{\rm Az}
\newcommand{\El}{\rm El}
\newcommand\urltilda{\kern -.15em\lower .7ex\hbox{\~{}}\kern -0.05em }
\newcommand{\Number}[1]{\ifnum#1<10\relax0\number#1\else\number#1\fi}
\newcommand{\isodate}{
\count151=\time
\divide\count151 by 60
\count151=\count151
\multiply\count151 by 60
\count152=\time
\advance\count152 by -\count151
\divide\count151 by 60
\count152=\count151
\multiply\count151 by 60
\count153=\time
\advance\count153 by -\count151
\Number{\year}.\Number{\month}.\Number{\day}--\Number{\count152}:\Number{\count153}
}

\begin{document}

\title{Precise geodesy with the Very Long Baseline Array}
\author{
        Leonid Petrov     \and
        David Gordon      \and
        John Gipson       \and
        Dan MacMillan     \and
        Chopo Ma          \and
        Ed Fomalont       \and
        R. Craig Walker   \and
        Claudia Carabajal
}
\institute{L. Petrov, \at
              ADNET Systems Inc./NASA GSFC, Code 610.2, Greenbelt, 
              MD 20771 USA \\
              \email{Leonid.Petrov@lpetrov.net} 
           \and
              D. Gordon, J. Gipson, D. MacMillan \at
              NVI Inc./NASA GSFC, Code 698, Greenbelt, MD 20771 USA \\
           \and
           C. Ma \at
              NASA GSFC, Code 698, Greenbelt, MD 20771 USA \\
           \and
           E. Fomalont, \at
              National Radio Astronomy Observatory, 
              520 Edgemont Rd, Charlottesville, VA~22903--2475, USA \\
           \and
           R.~C. Walker \at
              National Radio Astronomy Observatory, P.~O. 
              Box O, Socorro, NM 87801 USA.
           \and
           C. Carabajal \at Sigma Space Corporation/NASA GSFC, Code 698, 
              Greenbelt, MD 20771 USA 
           \and
           Published online: 28 February 2009
}

\date{Received: 01 June, 2008/ Accepted: 26 January 2009\\
$\copyright$ Springer-Verlag 2009}

\maketitle



\begin{abstract}
\par\vspace{-10mm}\par
   We report on a program of geodetic measurements between 1994 and 2007
which used the Very Long Baseline Array and up to 10 globally distributed
antennas. One of the goals of this program was to monitor positions of the 
array at a 1~millimeter level of accuracy and to tie the VLBA into the 
International Terrestrial Reference Frame. We describe the analysis of 
these data and report several interesting geophysical results including 
measured station displacements due to crustal motion, earthquakes, and 
antenna tilt.  In terms of both formal errors and observed scatter, these 
sessions are among the very best geodetic VLBI experiments.

\keywords{VLBI \and coordinate systems \and plate tectonics \and VLBA}
\end{abstract}

\section{Introduction} \label{s:intro}

   The method of very long baseline interferometry (VLBI), first proposed 
by \citet{r:mat65}, is a technique of computing the cross-power spectrum 
of a signal from radio sources digitally recorded at two or more 
radiotelescopes equipped with independent frequency generators. This spectrum 
is used in a variety of applications. One of the many ways of utilizing 
information in the cross-power spectrum is to derive a group interferometric 
delay \citep{r:ksp,r:tms}. It was shown by \citet{r:sha70} that group 
delays can be used for precise geodesy. The first dedicated geodetic 
experiment, on January 11, 1969, yielded 1~meter accuracy \citep{r:hin72}. 
In the following decades VLBI technology flourished, sensitivities and 
accuracies were improved by several orders of magnitude, and arrays of 
dedicated antennas were built. Currently, VLBI activities for geodetic 
applications are coordinated by the International VLBI Service for Geodesy 
and Astrometry (IVS) \citep{r:ivs}.

\begin{figure*}[ht]
  \ifprep
  {\includegraphics[width=\textwidth,clip]{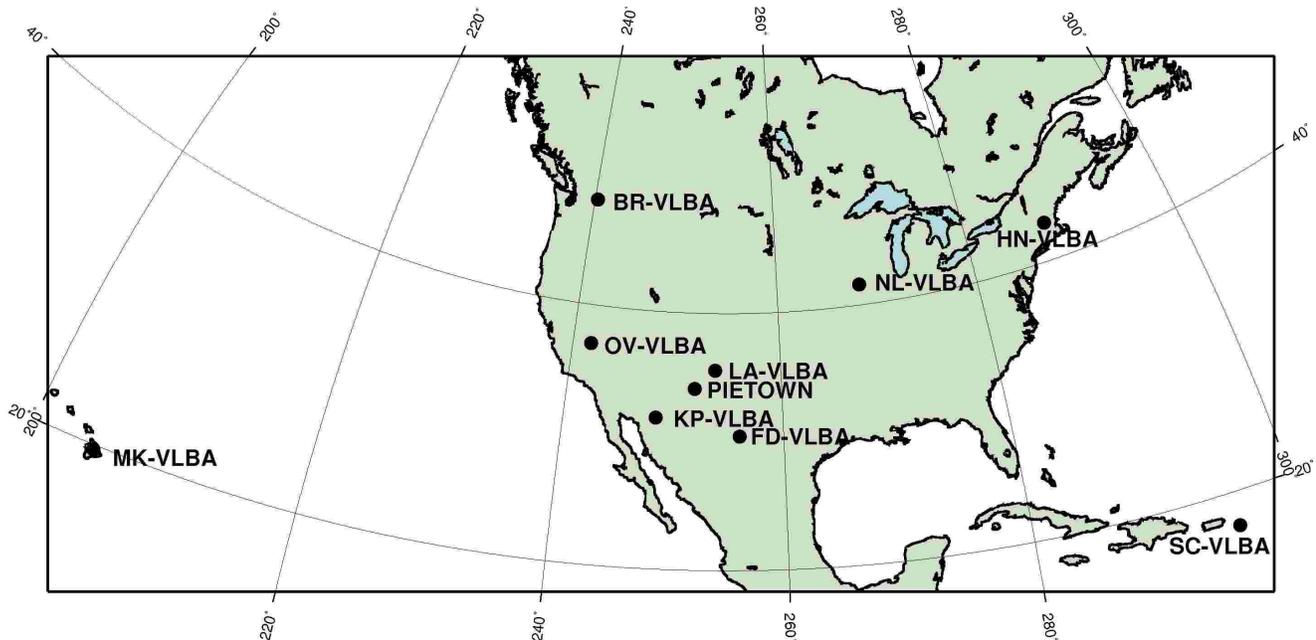} }
  \else
  {\includegraphics[width=\textwidth,clip]{vlba_map.eps} }
  \fi
  \par\vspace{-4ex}\par
  \caption{Positions of the antennas of the Very Long Baseline Array}
  \label{f:vlba_map}
\end{figure*}

\begin{figure}[hb]

  \ifprep
     \includegraphics[width=0.48\textwidth,clip]{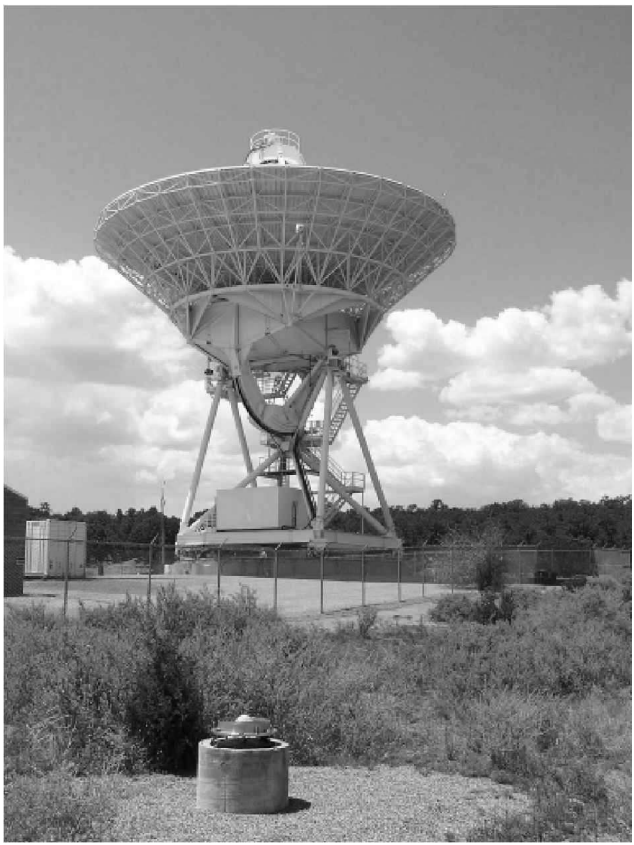}
  \else
     \includegraphics[width=0.48\textwidth,clip]{pietown_bw_lr.eps}
  \fi
  \caption{The VLBA station {\sc pietown} is in the background. 
           The permanent GPS receiver {\sc pie1} is in the foreground.}
  \label{f:pietown}
\end{figure}

  Among dedicated VLBI arrays, the Very Long Baseline Array (VLBA) 
\citep{r:n94} of ten 25~meter parabolic antennas spread over the US
territory (Figure~\ref{f:vlba_map}) is undoubtedly the most productive. 
The VLBA is a versatile instrument used primarily for astrometry and 
astrophysical applications. All ten VLBA antennas have identical design
(Figure~\ref{f:pietown}). They have an altitude-azimuth mounting with 
a nominal antenna axis offset of 2132~mm. Slewing rates are 
$1.5^\circ \: {\rm s}^{-1}$ in azimuth and $0.5^\circ \: {\rm s}^{-1}$ in 
elevation. Permanent GPS receivers are installed within 100 meters of 
5 antennas, {\sc br-vlba, mk-vlba, nl-vlba, pietown}, and {\sc sc-vlba}. 

  Phase referencing for detection of weak radio sources and for proper motion 
and parallax measurements are used in about half of all VLBA sessions.  
Accuracies on the order of 10~microarcseconds using source--to--calibrator 
separations of around one degree are achieved in the best current observations.
Such accuracies need to be supported by the underlying geometric model and 
its input parameters, including the station and source catalogs and the Earth 
orientation parameters (EOP). A future goal is to improve on this accuracy 
by a factor of 2 or more. To achieve 10~microarcsecond accuracy on a 4000~km 
baseline, a delay accuracy after calibration of 0.2~mm or 0.6~ps is required 
for any effects that cannot be reduced by integration.  Phase referencing over 
a one degree source--to--calibrator separation reduces model errors by a factor 
of 57, requiring the model parameters to be accurate to around 1 cm. Higher 
accuracies are desired to deal with the cumulative effect of several model 
parameters, to meet future goals, or to allow larger source--to--calibrator 
separations. 

  Use of the Global Positioning System (GPS) can provide very high
quality time series of site positions. Averaging these time series over 
several years can provide sub-mm estimates of the {\it phase center} 
positions of the GPS antennas, but this precision cannot be transferred to 
the {\it reference points} of the VLBI antennas for several reasons. First, 
measurements of the tie vector between the GPS phase center and the reference 
point of a radiotelescope introduce an additional uncertainty at a level 
of 3~mm or higher. Second, systematic errors of the GPS technique, such as 
phase center variations, multi-path, scale errors, and orbital errors, 
may cause biases in measurement of the phase center at a level of tens of mm. 
Third, a nearby GPS receiver may not experience the same localized effects
as the VLBI antenna, such as settling or tilting of the support
structure. According to \citet{r:ray2005} (Table~4 in their paper), the root 
mean square (rms) of differences between coordinates of VLBI reference points 
derived from analysis of VLBI observations and from analysis of GPS 
observations plus ties measurements among 25 pairs of GPS/VLBI sites are 
6~mm for the horizontal components and 13~mm for the vertical components 
after removal of the contribution of 14~Helmert transformation parameters 
fitted to the differences.

  The best way for determining positions of the antenna reference points 
is to derive them directly from dedicated geodetic VLBI observations on the 
VLBA array. Uncertainties of better than 1~mm are easily achieved.  Since 
the motion of these antenna reference points cannot be predicted precisely, 
geodetic observations need to be repeated on a regular basis in order 
to sustain that high precision.

  The importance of precise position monitoring was recognized during the 
design of the VLBA and each antenna began to participate in geodetic VLBI 
observations soon after it was commissioned. Between July 1994 and August 
2007, there were 132 dedicated 24 hour dual band S/X VLBI sessions 
under geodesy and absolute astrometry programs with a rate of 6--24 sessions 
per year. During each session, all ten VLBA antennas and up to 10 other 
geodetic VLBI stations participated. In this paper we present the geodetic 
results from this campaign. In section~\ref{s:setup} we describe the goals 
of the observations, scheduling strategies and the hardware configuration. 
In section \ref{s:correlation} we describe the algorithm for computing 
group delays from the output of the FX VLBA correlator and validation of 
the post-correlator analysis procedure. The results and the error analysis 
are presented in sections \ref{s:anal} and \ref{s:errors}. Concluding 
remarks are given in section \ref{s:conclusions}.
\section{Observing sessions} \label{s:setup}

  The primary goal of these geodetic VLBI observations was to derive an 
empirical mathematical model of the motions of the antenna reference points. 
The antenna reference point is the projection of antenna's moving axis 
(the elevation axis for altitude-azimuth mounts) to the fixed axis 
(the azimuthal axis for altitude-azimuth mounts). This mathematical model 
can be used for reduction of astronomical VLBA observations as well as for 
making inferences about the geophysical processes which cause this motion. 
A secondary goal was to estimate the precise absolute positions of many 
compact radio sources not previously observed under absolute astrometry 
programs, for use as phase referencing calibrators. Other goals, not 
discussed here, include monitoring a list of $\sim\! 400$ selected 
sources and producing time series of source coordinate estimates and images 
for improving the source position catalogue (A.~Fey et~al. (2009), paper 
in preparation) and for studying source structure 
changes \citep{r:rdv_astro_01,r:coreshift}.

  The observing sessions were typically 24 hours long. 
The radio sources observed were distant active galactic nuclei 
at distances of a gigaparsec scale\footnote{1~gigaparsec 
$\approx 3.2 \cdot 10^{9}$ light years $\approx 3 \cdot 10^{25}$ m} 
with continuum radio emission from regions of typically  
0.1--10~milliarcsecond in size.

  VLBA geodetic observations use the dual frequency S/X mode, 
observing simultaneously at S and  X~bands, centered around 2.3 and 8.6~GHz. 
This is enabled by a dichroic mirror permanently positioned over the S~band 
receiver, reflecting higher frequency radiation towards a deployable 
reflector leading to the X~band receiver. The system equivalent flux 
densities (SEFD) of VLBA antennas are in the range of 350--400~Jy when 
using the dual-frequency S/X system. From each receiver, four frequency 
channels 4~MHz wide before April~1995 and 8~MHz thereafter, were recorded 
over a large spanned bandwidth to provide precise measurements of group 
delays. The sequence of frequencies (called IF) was selected to minimize 
sidelobes in the delay resolution function and to reduce adverse effects 
of radio interference. The sequence was slightly adjusted over the 14~year 
period of observations in accordance with changes in the interference 
environment. The frequency sequence used in the session of 2007.08.01 
is presented in Table~\ref{f:frq}. 

\begin{table}[hb]
   \caption{The range of frequencies in the 
            observing session of 2007.08.01, in MHz.}
   \label{f:frq}
   \begin{flushleft}
      \begin{tabular}{lll}
         \hline \\
         IF1 & 2232.99 & 2240.99  \\
         IF2 & 2262.99 & 2270.99  \\
         IF3 & 2352.99 & 2360.99  \\
         IF4 & 2372.99 & 2380.99  \\
         IF5 & 8405.99 & 8413.99  \\
         IF6 & 8475.99 & 8483.99  \\
         IF7 & 8790.99 & 8798.99  \\
         IF8 & 8895.99 & 8903.99  \\
         \hline\vex 
      \end{tabular}
   \end{flushleft}
   \par\vspace{-14ex}\par
\end{table}

\subsection{Scheduling}

  Among the 132 observing sessions, 97 can be characterized as global geodetic 
sessions and 35 as absolute astrometry sessions. They
differ in scheduling strategy. A wider list of 150--250 sources was observed
in each astrometry session while a shorter list of $\sim\!\! 100$ objects was 
observed in each geodesy session. 

\subsubsection{Scheduling of astrometric sessions}

  Two lists of sources were observed in astrometry sessions: a list of 
150--200 target sources and a list of 30--80 tropospheric calibrators. 
Selection of tropospheric calibrators was based on two 
criteria: a)~the compactness at both X and S band, i.e. the ratio of the 
median correlated flux density at baselines longer than 5000~km to the 
median correlated flux density at baselines shorter than 900~km, must be 
greater than 0.5; b)~the correlated flux density at baselines longer than 
5000~km must be greater than 0.4~Jy at both X and S~bands. These sources 
are frequently observed in other IVS geodetic programs.

  Target sources were scheduled for 1--3 scans, i.e. the period of time when 
antennas are on source and record the data, in a sequence that seeks 
to minimize slewing time needed for pointing all antennas to the next source. 
In the astrometry sessions, normally all antennas simultaneously observe 
the same object for the same duration. Scan durations were determined on 
the basis of the predicted correlated flux density and the SEFDs to get  
SNRs of the multi-band fringe amplitude greater than 20. The typical scan 
durations were 40--480~s. 

  The sequence of target sources was interrupted every 1.5 hours, to 
observe 3--5 tropospheric calibrators. The tropospheric calibrators were 
scheduled in such a way that at each station, at least one calibrator was 
observed in the ranges of $[7^\circ, 20^\circ]$ elevation, 
$[20^\circ, 50^\circ]$  elevation, and above $50^\circ$ elevation. 
The purpose of including tropospheric calibrators was a)~to reliably estimate 
the zenith path delay of the neutral atmosphere in the least squares (LSQ) 
solution, and b)~to link the positions of new or rarely observed target 
sources with those of frequently observed calibrators. Astrometric schedules 
were prepared with the NRAO software package SCHED. The efficiency of these 
schedules, i.e. the ratio of time on source to the total time of the 
observing session is typically $\sim\! 70\%$.

\subsubsection{Scheduling of geodetic sessions}

  The geodetic sessions involved $\sim\! $15--20 geographically dispersed 
antennas with varying sensitivities. At any given time, few sources, if any,
are visible by the entire network. Hence, in contrast to the astrometric 
sessions, at any instant different subsets of antennas will be observing 
different sources, and the integration time will vary from antenna to antenna 
in order to reach the required SNR. The minimum elevation angle for scheduled 
observations for all antennas in the geodetic VLBA sessions is set 
to $5$~degrees.

  These sessions were scheduled using the automatic scheduling mode of 
the SKED program. The scheduler sets up  some general parameters that govern 
how the schedule is generated.  The scheduler then generates all or part of 
the schedule and examines it for problems, such as prolonged gaps in the 
schedule when a station is idle. The scheduling parameters can be adjusted 
to minimize problems.  The schedule can also be modified by adding or 
deleting observations. In its automatic mode, SKED generates a sequence 
of scans using the following algorithm:

\begin{enumerate}

    \item SKED determines the current schedule time by looking at the
          latest time any station was scheduled, and taking the earliest 
          of these times. 

    \item SKED updates the logical source-station visibility table for the 
          current time. The rows of this table correspond to sources and the 
          columns to stations. If a source is visible at a station, the location 
          is marked as true, else it is false. Any row that has two or more true 
          entries corresponds to a possible scan.This table is modified by the 
          so-called ``Major Options'' which control which scans are actually 
          considered. The important major options are:  A)~{\tt MinBetween}.  
          If a source has been observed more recently than {\tt MinBetween}, 
          it is marked as down at all stations.  This prevents strong sources 
          from being observed too frequently. B)~{\tt MaxSlew}. If the slew 
          time for an antenna is longer than
          {\tt MaxSlew}, the source is marked as down for the station. 
          C)~{\tt MinSubNetSize}. If the number of stations which can see 
          a given source is smaller than {\tt MinSubNetSize}, the source 
          is marked as down at all stations.

    \item SKED scores scans based on their effect on {\tt Sky Coverage} or
          {\tt Covariance} optimizations.  The user has the option of 
          choosing which, with {\tt Sky Coverage} the usual choice.  
          A)~For {\tt Sky Coverage}, SKED calculates, 
          for each station, the angular distance of the source from 
          all previous scans over some time interval.  It finds the 
          minimum angular distance, and averages over all stations. 
          This is the sources' sky-coverage score.  The larger the score, 
          the larger the hole that will be filled by observing this source. 
          B)~For {\tt Covariance} optimization, prior to scheduling, the 
          scheduler specifies a set of parameters to be estimated, and 
          a subset to optimize.  For example, you might estimate 
          atmosphere at each station, clocks at all but the reference 
          station, and EOP, but you are only interested in optimizing EOP. 
          Scans are ranked by the decrease in the sum of the formal errors
          of the optimized parameters when the considered scan is added 
          to the schedule. C)~In either case, the top {\rm X\%} of scans 
          are kept for further consideration, where  {\rm X\%} is user 
          settable, and is typically 30--50\%. The smaller {\rm X\%} is, 
          the more important the initial ranking.  

    \item Lastly the top {\rm X\%} scans are ranked by a set of 
          ``Minor Options''. There are 15 Minor Options, each 
          corresponding to some possible desirable feature of the scan. 
          For each scan, SKED calculates the weighted sum of the minor 
          options in use. The scan with the highest overall score is 
          scheduled.  A description of all of the Minor Options and how 
          the score is calculated is beyond the scope of this paper. The
          Minor Options typically used for scheduling geodetic VLBA 
          experiments, and their effect on the scan selection follows. 
          A)~{\tt EndScan} prefers scans which end soonest. B)~{\tt NumObs}  
          prefers scans with more observations, i.e., with more stations. 
          C)~{\tt StatWt} prefers scans involving certain stations. This 
          is a way of increasing the number of observations at weak stations, 
          or stations that are poorly connected to the network.  
          D)~{\tt StatIdle} prefers scans which involve stations which have 
          been idle. This reduces gaps in the schedule. 
          E)~{\tt Astrometric} and F)~{\tt SrcEvn} modes are discussed 
          below.

\end{enumerate}

  When SKED is done scheduling a scan, it checks to see if there is more
time left, in which case it returns to Step 1.  If not, it returns
control to the scheduler. 

  Geodetic VLBA experiments have two goals absent from other geodetic 
VLBI sessions: 1)~The inclusion of ``requested'' sources for which precise 
positions have been requested by the astronomical community;  and 2)~The 
desire to image all (or most) of the sources in each experiment. These lead 
to the development of {\tt Astrometric} mode and {\tt SrcEvn} modes in SKED. 
In {\tt Astrometric} mode the user specifies minimum and maximum 
observing targets for a list of sources. SKED preferentially selects scans 
involving sources which are below their targets, and discriminates against 
scans involving sources which are above their targets. {\tt SrcEvn} mode was 
introduced because SKED has a tendency to select strong sources with 
good mutual visibility. If {\tt SrcEvn} mode is turned on, SKED will 
preferentially schedule sources that are under-observed compared to their 
peers. This is one way of ensuring that weak sources, or sources with low 
mutual visibility, are observed a sufficient number of times so that they 
can be imaged. The efficiency of geodetic schedules is typically 45--60\%.

\subsection{Session statistics}

  The distribution of sessions over time is presented in 
Table~\ref{t:statexp}. In each session 7,000--34,000 pairs of S/X group 
delays were evaluated, for a total of 1,737\,947 values. The ten VLBA 
stations and 20 other non-VLBA stations took part in the observing campaign,
with from 9 to 20 stations in each session.  The frequency of station 
participation in sessions is shown in Table~\ref{t:sitexp}. Among 4412 
observed sources, at least two usable S/X pairs of group delays were 
determined for 3090 objects.

\begin{table}[ht]
   \caption{Statistics of VLBA sessions 
   }
   \label{t:statexp}
   \begin{flushleft}
      \begin{tabular}{lrr}
         \hline \vspace{-2ex}\\
          \nc{Year} & \# geodetic & \# astrometric      \\
                    & sessions    &  sessions      \vex \\
         \hline
            1994  &      3 \hps   &    1 \hps \\
            1995  &     12 \hps   &    2 \hps \\
            1996  &     16 \hps   &    8 \hps \\
            1997  &      6 \hps   &    5 \hps \\
            1998  &      6 \hps   &    0 \hps \\
            1999  &      6 \hps   &    0 \hps \\
            2000  &      6 \hps   &    0 \hps \\
            2001  &      6 \hps   &    0 \hps \\
            2002  &      6 \hps   &    2 \hps \\
            2003  &      6 \hps   &    0 \hps \\
            2004  &      6 \hps   &    4 \hps \\
            2005  &      6 \hps   &    7 \hps \\
            2006  &      7 \hps   &    5 \hps \\
            2007  &      5 \hps   &    1 \hps \\
         \hline\vex 
      \end{tabular}
   \end{flushleft}
   \par\vspace{-10ex}\par
\end{table}

\begin{table}[ht]
   \caption{Statistics of observing session per station.
            (1) IVS station name; 
            (2) geocentric latitude; 
            (3) longitude, positive towards east; 
            (4) Number of observing sessions under geodesy and astrometry 
                with the VLBA array.}
   \label{t:sitexp}
   \begin{flushleft}
      \begin{tabular}{l@{\qquad}r@{\qquad}r@{\qquad}r}
         \hline \vspace{-2ex}\\
            \nc{(1)}       & \nc{(2)} & \nc{(3)}  & \nc{(4)} \vex \\
         \hline
            \sc{pietown }  & $+$34.122 &  251.880   &  132  \\
            \sc{la-vlba }  & $+$35.592 &  253.754   &  130  \\
            \sc{kp-vlba }  & $+$31.783 &  248.387   &  129  \\
            \sc{br-vlba }  & $+$47.939 &  240.316   &  128  \\
            \sc{ov-vlba }  & $+$37.046 &  241.722   &  128  \\
            \sc{fd-vlba }  & $+$30.466 &  256.055   &  126  \\
            \sc{hn-vlba }  & $+$42.741 &  288.013   &  126  \\
            \sc{mk-vlba }  & $+$19.679 &  204.544   &  124  \\
            \sc{nl-vlba }  & $+$41.580 &  268.425   &  123  \\
            \sc{sc-vlba }  & $+$17.645 &  295.416   &  117  \\
            \sc{westford}  & $+$42.431 &  288.511   &   59  \\
            \sc{kokee   }  & $+$21.992 &  200.334   &   57  \\
            \sc{gilcreek}  & $+$64.830 &  212.502   &   56  \\
            \sc{onsala60}  & $+$57.220 &   11.925   &   50  \\
            \sc{wettzell}  & $+$48.954 &   12.877   &   50  \\
            \sc{medicina}  & $+$44.328 &   11.646   &   46  \\
            \sc{nyales20}  & $+$78.856 &   11.869   &   38  \\
            \sc{tsukub32}  & $+$35.922 &  140.087   &   31  \\
            \sc{hartrao }  & $-$25.738 &   27.685   &   28  \\
            \sc{ggao7108}  & $+$38.833 &  283.173   &   24  \\
            \sc{tigoconc}  & $-$36.658 &  286.974   &   21  \\
            \sc{nrao20  }  & $+$38.245 &  280.160   &   20  \\
            \sc{matera  }  & $+$40.459 &   16.704   &   19  \\
            \sc{hobart26}  & $-$42.611 &  147.440   &    6  \\
            \sc{kashim34}  & $+$35.772 &  140.657   &    6  \\
            \sc{algopark}  & $+$45.763 &  281.927   &    5  \\
            \sc{noto    }  & $+$36.691 &   14.989   &    4  \\
            \sc{zelenchk}  & $+$43.595 &   41.565   &    3  \\
            \sc{svetloe }  & $+$60.367 &   29.781   &    2  \\
            \sc{urumqi  }  & $+$43.279 &   87.178   &    1  \\
         \hline\vex 
      \end{tabular}
   \end{flushleft}
   \par\vspace{-8ex}\par
\end{table}

\section{Correlation and post-correlation analysis of observations}
\label{s:correlation}

  Observations at individual stations were recorded on magnetic tapes or,
since 2007, on Mark~5 disc packs. Cross-correlation of the raw data was 
performed on the VLBA correlator \citep{r:corr_benson,r:corr_walker}, 
in Socorro, N.M., USA. The correlator uses the GSFC program Calc and the 
station clock offsets with respect to UTC measured with GPS clocks to compute 
theoretical delays to each station. Each station's bit stream is offset 
by these delays during the correlation. The resultant correlator output 
is the amplitudes and residual phases as functions of time (visibility 
points) for each station, referenced to a common point that lies close to, 
but not necessarily coincident with the geocenter. 

  Most geodetic VLBI experiments are correlated using Mark~4 correlators
\citep{r:mark-4}. Their output is processed using the Fourfit program 
developed at MIT Haystack Observatory. Since this program cannot 
handle the output from the FX correlators, we used the AIPS software 
package \citep{r:aips} for further processing.

\subsection{Using AIPS to process geodesy experiments}

  Additional processing is required to evaluate the geodetic/astrometric VLBI 
observables of group delays and phase delay rates. The initial calibrations are:

\begin{enumerate}

  \item Small amplitude corrections for the correlator statistics are 
        applied while reading the raw data into an AIPS data base.

  \item The reference point of each IF channel is moved from the 
        lower frequency edge to the center frequency of the channel,
        along with an adjustment to the frequency in the AIPS data base.
        This reduces edge effects, resulting in a small improvement
        in determination of the group delays. 
%

  \item Bad antenna and frequency channels are flagged out, as necessary.

 \item  At the VLBA stations, relative phase and delay offsets are applied to 
        the visibility points using measured phase calibration tone phases. 
        For both VLBA and non-VLBA stations, manual phase offsets 
        are applied.\footnote{The non-VLBA stations have phase calibration 
        systems, but their phases could not be captured in real time, nor 
        extracted during correlation as is done on the 
        Mark~4 correlators.}  The phase offsets are determined by fringe fitting
        a reference scan on a compact radio source to determine the relative 
        instrumental phase and residual group delay for each individual IF. 
        These phases are removed from the entire data set, equivalent 
        to setting the single band and multiband residual delays to zero 
        at the scan used for the calibration.

\end {enumerate}

  The heart of the reduction process is the {\it fringe fitting} of the
data using AIPS task FRING. Data for each scan, baseline, frequency band and 
IF channel are processed separately, and the following parameters are 
determined: the average phase at some fiducial time near the center of the 
scan; the average phase rate with time (fringe rate); and the average phase 
rate with frequency (single-band delay) by finding the maximum
of the 2D Fourier-transform of visibility data \citep{r:ksp} and 
subsequent LSQ fit. An SNR cutoff of about 3 is generally used in order 
to omit noisy solutions for relatively weak sources\footnote{The AIPS cookbook 
can be found on the Web at {\tt http://www.aips.nrao.edu/cook.html}}.

  For those observations in which all of the IF channels have a detection, 
an AIPS program called MBDLY computes the average phase for the reference 
frequency and the average phase slope with frequency (so-called multi-band 
delay or group delay) that best fits the individual IF phases obtained from 
FRING. The individual IF solutions for the single-band delay and the fringe 
rate are averaged over all IFs. Checks of the quality of the group delays 
are obtained by the spread of the individual single-band delays, the fringe 
rates, and the phase scatter between each measured IF phase versus the 
best-fit group delay. Observations with large deviations are flagged as low 
quality and generally are not used in the analysis. 

  The results from FRING and MBDLY give the phase, single-band delay,
fringe-rate, and group delay for a fiducial time near the middle of
each scan for each baseline and each frequency band. These quantities represent 
the residual values with respect to the correlator model for the observation.  
When these data are added to that of the correlator model, the results become 
the {\it total} phase delay, {\it total} single-band delay, {\it total} 
fringe-rate and {\it total} group delay, respectively. These total values are 
independent of the correlator model.

  The correlator model is attached to the correlator output. For each source 
and each antenna, it is represented by a six-order polynomial for every two 
minute interval, so that its value can be determined at any time with 
rounding errors below 0.1~ps. It contains three parts: the geometric delay 
based on the a~priori source position, antenna locations, and the Earth 
orientation parameters; an a~priori atmospheric delay; and the clock offset 
with respect to UTC determined by GPS at each individual station. The time-tag
associated with the correlator model and the residual parameters is 
earth-center oriented. That is, the parameters are referenced to the time 
when the wavefront intercepts a fiducial point chosen at the coordinate 
system origin in order to facilitate the correlation process. The total 
quantities are then determined by adding the baseline residual parameters 
to the correlator model difference for the appropriate antenna-pair, 
interpolated to the scan reference time.

  The total observables are continuous functions of time. Further geodetic 
and astrometric analysis requires discrete values of observable quantities,
one per scan and per baseline, thereafter called observations, with time-tags 
associated with the arrival of the wavefront at the reference antenna of 
the baseline. These observables can differ by as much as 20~ms from the 
quantities with Earth-centered time-tags. AIPS task CL2HF is used to combine 
the correlator models and residuals at two stations and compute the 
observables with the reference antenna time-tags. For convenience, the 
time-tags are chosen to be on an integer second, and a common time-tag 
is set to all observations in a scan. CL2HF performs this transformation,
computes the fringe amplitude SNRs, delay, and rate uncertainties. 
CL2HF writes out an ``HF'' extension AIPS file which contains the total 
quantities as well as many other derived quantities needed for further
analysis. Finally, the AIPS task HF2SV converts the data in the HF extension 
file to a binary form that is consistent with Mark~3 correlator output.

\subsection{Validation of the post-correlation analysis procedure}
\label{s:validation}

  For the first few years, the VLBA/AIPS processed sessions were freely mixed 
with Mark~4/Fourfit processed sessions, with few noticeable effects. 
However, two discrepancies were noticed between results from the two data 
sets. 1)~The horizontal position of the {\sc onsala} station shifted by 
approximately 3~mm between the two sets of data and 2)~scatter of source 
position series for southern sources differed at a level of 0.2--1.0~mas. 
The shift in {\sc onsala}'s position was found to be the result of a strong 
azimuthal dependence of instrumental delay in the cable, not seen at other 
sites. It showed up because measured phase calibration was not used 
for {\sc onsala} in the VLBA/AIPS processing. The source statistics 
difference was found to be due to an incorrect accounting in program CL2HF 
of the total number of bits read from the station tapes at the VLBA 
correlator. When this was corrected, the ``southern source'' problem 
disappeared. 

A direct comparison of delays and rates processed by the AIPS software package
versus the Haystack Fourfit software package was strongly desired. To make 
such a comparison, the tapes from 8 stations in the rdv22 VLBA session 
(2000 July 6--7) were saved and sent to Haystack Observatory, where they were 
correlated on the Mark~4 correlator and fringed using the program Fourfit. 
To minimize the differences in processing, a single set of phase 
calibration phases was used in both the AIPS and Fourfit processing. 
Two databases were made of the same baselines processed through the two 
independent systems, with matching time tags. The regular Calc/Solve analysis 
was then performed on each to eliminate any bad data points. The observed 
delays and rates were differenced and tabulated by baseline. There were 
constant offsets for each baseline due to differences in single band 
calibration, and differences in $2\pi$ ambiguity shifts. Such delay 
differences get absorbed into the clock adjustments and do not affect the 
geodetic or astrometric results. After removal of these constant differences, 
weighted root mean square (wrms) differences at X-band were computed by 
baseline. These are given in Table~\ref{t:rdv22}. The wrms differences range 
from as little as 2.5~ps on the {\sc kp-vlba/ov-vlba} baseline, up to 
16.1~ps on the {\sc br-vlba/mk-vlba} baseline. 
The wrms over all baselines is only 6.0~ps, equivalent to 1.8~mm. 
By comparison, the average delay formal errors are 7.7~ps on 
{\sc kp-vlba/ov-vlba} and 17.3~ps on {\sc br-vlba/mk-vlba}. 
In a similar comparison between the Mark~3 and Mark~4 correlators and 
post-processing software at Bonn University, \citet{r:n2002} found an 
average wrms difference of 21.1~ps on 6 long intercontinental baselines.

\begin{table}
   \caption{A comparison of X-band group delays and phase delay rates 
            between a subset of the rdv22 session processed through the 
            Haystack Mark~4 correlator/Fourfit processing system versus 
            the VLBA Correlator/AIPS processing system.}
   \label{t:rdv22}
   \begin{tabular}{lrrrrr}
       \hline \\
                 &         &    & \nntab{c}{WRMS Differences} \\
       Baseline  & \#pts   & Length     & Delays       & Rates \\
                 &         & (km)       & (ps)         & $10^{-15}$ \\
       \hline
        \sc{la-vlba }/\sc{pietown }  & 135  &  237  &   5.1  & 31.3   \\
        \sc{kp-vlba }/\sc{pietown }  & 124  &  417  &   3.6  & 14.1   \\
        \sc{kokee   }/\sc{mk-vlba }  & 169  &  508  &   5.1  & 35.1   \\
        \sc{kp-vlba }/\sc{la-vlba }  & 125  &  652  &   5.2  & 15.8   \\
        \sc{kp-vlba }/\sc{ov-vlba }  & 171  &  845  &   2.5  & 61.1   \\
        \sc{ov-vlba }/\sc{pietown }  &  97  &  973  &   7.4  & 30.0   \\
        \sc{la-vlba }/\sc{ov-vlba }  &  76  & 1088  &   5.6  & 22.1   \\
        \sc{br-vlba }/\sc{ov-vlba }  & 176  & 1215  &   3.3  & 30.4   \\
        \sc{br-vlba }/\sc{la-vlba }  & 104  & 1757  &   3.0  & 30.3   \\
        \sc{br-vlba }/\sc{pietown }  & 170  & 1806  &   7.8  & 60.4   \\
        \sc{br-vlba }/\sc{kp-vlba }  & 183  & 1914  &   6.8  & 20.8   \\
        \sc{br-vlba }/\sc{gilcreek}  & 143  & 2482  &   4.0  & 10.2   \\
        \sc{gilcreek}/\sc{ov-vlba }  &  24  & 3584  &   6.1  & 51.9   \\
        \sc{mk-vlba }/\sc{ov-vlba }  & 124  & 4015  &   7.4  &  7.4   \\
        \sc{kokee   }/\sc{ov-vlba }  &  59  & 4220  &   8.4  & 36.8   \\
        \sc{gilcreek}/\sc{pietown }  & 103  & 4225  &   4.8  & 43.3   \\
        \sc{gilcreek}/\sc{kp-vlba }  &  98  & 4322  &   3.8  &  9.4   \\
        \sc{br-vlba }/\sc{mk-vlba }  & 162  & 4399  &  16.1  & 23.2   \\
        \sc{kp-vlba }/\sc{mk-vlba }  &  43  & 4467  &  13.3  & 33.2   \\
        \sc{br-vlba }/\sc{kokee   }  & 112  & 4469  &   9.2  & 11.3   \\
        \sc{gilcreek}/\sc{kokee   }  &  95  & 4728  &  15.4  & 36.5   \\
        \sc{kokee   }/\sc{kp-vlba }  & 113  & 4736  &   5.6  & 24.9   \\
        \sc{mk-vlba }/\sc{pietown }  &  38  & 4796  &   9.0  & 39.7   \\
        \sc{gilcreek}/\sc{mk-vlba }  &  21  & 4923  &  11.6  & 53.8   \\
        \sc{la-vlba }/\sc{mk-vlba }  &  89  & 4970  &   9.3  & 51.1   \\
        \sc{kokee   }/\sc{pietown }  & 105  & 5040  &   7.9  & 77.2   \\
        \\
        ALL                          & 2859 & \rm{---} &  6.0  & 36.1 \\
       \hline \\
   \end{tabular}
   \par\vspace{-5.5ex}\par
\end{table}

\section{Geodetic analysis}   \label{s:anal}

  In our analysis we used {\it all available} VLBI observations from 
August~03, 1979 through October~04, 2007, including 132 observing sessions 
with the VLBA. The differences between the observed ionosphere-free linear 
combinations of dual-frequency group delays and theoretical group delays 
are used in the right hand side of the observation equations in the least 
squares parameter estimation procedure.

  Computation of theoretical time delays in general follows the approach 
outlined in the IERS Conventions \citep{r:iers03} and presented in detail by 
\citet{r:masterfit} with some minor refinements. The most significant ones
are the following. The expression for time delay derived by \citet{r:Kop99} in 
the framework of general relativity was used. The displacements caused by 
the Earth's tides were computed using a rigorous algorithm \citep{r:harpos} 
with a truncation at a level of 0.05~mm using the numerical values of the 
generalized Love numbers presented by \citet{r:mat01}. The displacements 
caused by ocean loading were computed by convolving the Greens' functions 
with ocean tide models using the NLOADF algorithm of \citet{r:spotl2}. 
The GOT00 model \citep{r:got99} of diurnal and semi-diurnal ocean tides, 
the NAO99 model \citep{r:nao99} of ocean zonal tides, the equilibrium model 
\citep{r:harpos} of the pole tide, and the tide with period of 18.6 years 
were used. Atmospheric pressure loading was computed by convolving the 
Greens' functions with the output of the atmosphere NCEP Reanalysis 
numerical model \citep{r:ncep}. The algorithm of computations is described 
in full details in \citet{r:aplo}. The empirical model of harmonic variations
in the Earth orientation parameters {\tt heo\_20070802} derived from VLBI 
observations according to the method proposed by \citet{r:erm} was used. 
The time series of UT1 and polar motion derived by the Goddard operational
VLBI solutions were used as a~priori. Displacement of the VLBI reference points
due to antenna thermal expansion was not modeled.

  The ionosphere contribution to group delay is considered to be reciprocal 
to the square of frequency. Therefore, there exists the linear combination of 
X-band and S-band delays that is ionosphere-free. No additional ionosphere model 
was applied. The contribution of higher terms to the ionosphere delay as was 
shown by \citet{r:iono2nd} is less than 9~ps. Its maximum contribution 
to estimates of site positions is below 0.5~mm and it was ignored.

  The a~priori path delay in the atmosphere caused by the hydrostatic 
component was calculated as a product of the zenith path delay computed
on the basis of surface pressure using the \citet{r:Saa72a,r:Saa72b} 
expression with corrections introduced by \citet{r:davis} and the so-called 
hydrostatic mapping function \citep{r:nmf}. The mapping function describes 
the dependence of path delay on the angle between the local axis of symmetry 
of the atmosphere and the direction to the observed sources. 

   Several solutions were produced. Each solution used the basic 
parameterization which was common for all runs and a specific parameterization
for an individual solution. Basic parameters belong to one of the 
three groups: 
\begin{itemize}
       \item [---]{\it global} (over the entire data set): positions of 
                          3089 sources.

       \item [---]{\it local}  (over each session):  
                          tilts of the local symmetric axis of the atmosphere
                          (also known as ``atmospheric azimuthal gradients'') 
                          for all stations and their rates, station-dependent 
                          clock functions modeled by second order polynomials, 
                          baseline-dependent clock offsets, daily nutation 
                          offset angles.

       \item [---]{\it segmented} (over 20--60 minutes): coefficients of 
                          linear spline that model atmospheric path delay
                          (20 minutes segment) and clock function 
                          (60 minutes segment) for each station. The estimates 
                          of clock function absorb uncalibrated instrumental
                          delays in the data acquisition system.
\end{itemize}

  The rate of change for the atmospheric path delay and clock function between 
adjacent segments was constrained to zero with weights reciprocal to 
$ 1.1 \cdot 10^{-14} $ and \mbox{$2\cdot10^{-14}$}, respectively, in order 
to stabilize solutions. The weights of observables were computed as
$ w = 1/\sqrt{\sigma_o^2 + r^2(b)} $, where $\sigma_o$ is the formal 
uncertainty of group delay estimation and $r(b)$ is the baseline-dependent
reweighting parameter that was evaluated in a trial solution to make the
ratio of the weighted sum of the squares of residuals to its mathematical 
expectation to be close to unity.

\subsection{Baseline analysis} \label{s:bas_anal}

  In the preliminary stage of data analysis, in addition to basic parameters
we estimated the length of each baseline at each session individually. 
The purpose of this solution was to determine possible non-linearity in
station motion, to detect possible outliers, and to evaluate statistics related 
to systematic errors. The baseline length is invariant with respect to
a linear coordinate transformation that affects all the stations of the 
network. Therefore, changes in baseline lengths are related to either
physical motion of one station  with respect to another or to systematic
errors specific to observations at the stations of the baseline.

\begin{figure}[htb]
  \label{f:baslen_hnfd}
  \vex\vex
  \includegraphics[width=0.48\textwidth,clip]{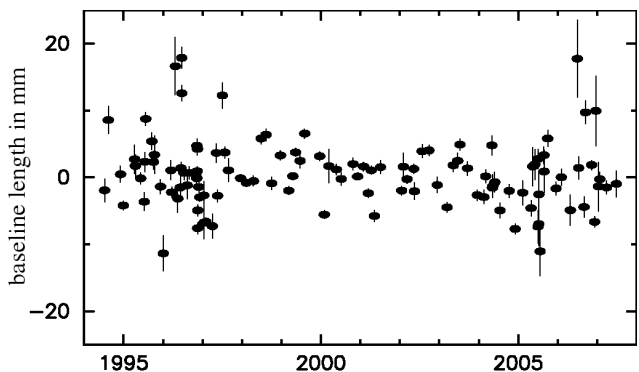}
  \caption{Residual lengths of the intra-plate baseline {\sc hn-vlba/fd-vlba} 
           with respect to the average value of $3\,623\,021.2526$~m. 
           The wrms 3.7~mm.}
  \par\vspace{-8ex}\par
\end{figure}

\begin{figure}[htb]
  \label{f:baslen_mksc}
  \vex\vex
  \includegraphics[width=0.48\textwidth,clip]{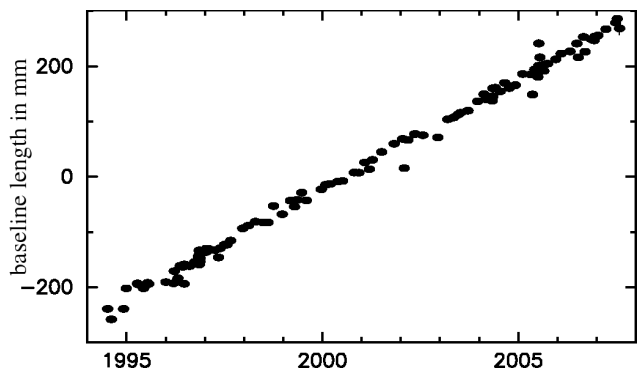}
  \caption{Residual lengths of the inter-plate baseline {\sc mk-vlba/sc-vlba} 
           with respect to the average value of $8\,611\,584.6972$~m.
           The wrms 9.2~mm.}
  \par\vspace{-2ex}\par
\end{figure}

  We present in Figures~\ref{f:baslen_hnfd}--\ref{f:baslen_mksc} examples 
of length evolutions for a very stable intra-plate baseline and for a rapidly 
stretching inter-plate baseline.

  As we see, the tectonic motion has shifted station {\sc mk-vlba}, located 
on the fast Pacific plate, by more than 0.5~meters over the existence 
of the array.

\begin{figure}[htb]
  \includegraphics[width=0.48\textwidth,clip]{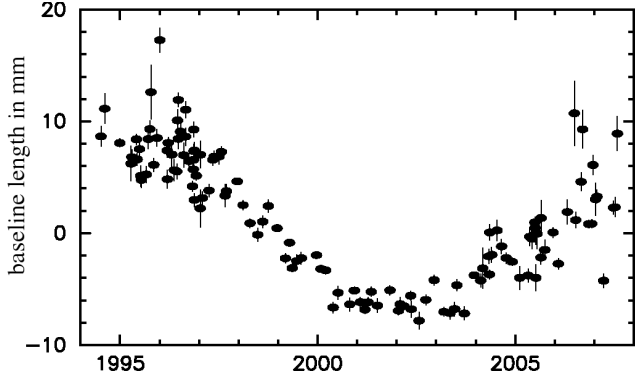}
  \caption{Length of the inter-plate baseline {\sc kp-vlba/pietown} with 
           respect to the average value of $417\,009.1252$~m.
           The baseline length estimates exhibit a non-linear motion caused 
           by variations in the antenna's tilt.}
  \label{f:baslen_kppt}
\end{figure}

  No significant outliers, and no jumps exceeding 2~cm were found in
examining plots of the baseline length evolution. Significant non-linear 
motion was found only on baselines with station {\sc pietown} 
(Figure~\ref{f:baslen_kppt}). Analysis of GPS data from the 
permanent IGS station PIE1 located within 61.8~m of the VLBI station 
(Figure~\ref{f:pietown}) does not show a similar pattern.

\subsection{Global analysis}   \label{s:glo_anal}

  The purpose of the global solution is to determine the best model of station
motion. In general, the model of motion of the $k$th station can be 
represented in this form:

\par\vspace{-2ex}\par
\begin{eqnarray} 
   \begin{array}{ll}
     \vec{r}_k = & \vec{r}_{ok} \: + \: \vec{\dot{r}}_k \, t\: + \!\!\!
                   \dss\sum_{j=1-m}^{n_k} \vec{f}_{kj} \, 
                   B^m_j(t;t_{1-m,k}, \ldots t_{n_k,k} ) \: + \: \\   &
                   \dss\sum_i^{n_h} \bigl(
                         \vec{h}^c_{ki} \cos ( \alpha_i + \omega_i t ) + 
                         \vec{h}^s_{ki} \sin ( \alpha_i + \omega_i t ) 
                                \big).
   \end{array}
   \label{e:ga1}
\end{eqnarray} 

  Here $ \vec{r}_{ok} $ is the position of the $k$th station at the reference 
epoch when t=0, $ \vec{\dot{r}}_k $ is the linear station velocity, 
$ B^m_j(t;t_{1-m,}, \ldots, t_{n_k,k}) $ is the B-spline of $m$th degree 
defined on a knot sequence \mbox{$t_{1-m,k}, \ldots, t_{n_k,k}$} that is unique
for each station and not necessarily equidistant with the $j$th pivotal 
element. Properties of B-spline function are discussed in full
details in \citet{r:boor_spl,r:nur_spl}. The first two terms in~\ref{e:ga1} 
describe the linear station motion, the last one describes harmonic motion, 
and the third term describes the non-linear, anharmonic motion with possible 
discontinuities caused by seismic events or antenna repair. 

  The parameters of the non-linear model of motion for selected sites, 
the frequencies of harmonic site position variations, the degree of 
the B-splines, and the sequences of knots on which the B-splines are defined, 
were selected manually. Several trial solutions were made, and the series
of the baseline length estimates were scrutinized. The parameters of the 
non-linear model were adjusted until the plots of residuals showed no 
systematics. The stations for estimation of harmonic position 
variations were selected on the basis of their observational history. 
Only those stations that participated in observations at least once every
three months for at least three years were selected to avoid strong 
correlation between estimates of harmonic site position variations and 
other parameters.


  We estimated non-linear anharmonic motion at 18 stations, including  
two VLBA stations {\sc pietown} and {\sc mk-vlba}. The degree of \mbox{B-spline} 
was 0 for {\sc mk-vlba} and 2 for {\sc pietown}. The epochs of B-spline knots 
are presented in Table~\ref{t:spline}.

\begin{table}[h]
   \caption{Epochs of knots of B-spline for modeling a non-linear anharmonic 
            motion of two VLBA stations.}
   \label{t:spline}
   \begin{tabular}{ll @{\qquad\qquad\qquad} ll}
        \hline
        \sc{pietown } & 1988.09.08 & \sc{mk-vlba } & 1993.07.19 \\
        \sc{pietown } & 1993.03.01 & \sc{mk-vlba } & 2000.04.02 \\
        \sc{pietown } & 1996.01.01 & \sc{mk-vlba } & 2006.10.15 \\ 
        \sc{pietown } & 1998.01.01 & \sc{mk-vlba } & 2007.08.10 \\ 
        \sc{pietown } & 2000.01.01 &                            \\
        \sc{pietown } & 2002.01.01 &                            \\
        \sc{pietown } & 2004.01.01 &                            \\
        \sc{pietown } & 2007.08.10 &                            \\
        \hline
   \end{tabular}
   \par\vspace{-3ex}\par
\end{table}

  We ran a special global solution\footnote{Listing of this solution is 
available at \hphantom{MMMMMMMMMM} 
{\tt http://astrogeo.org/vlbi/solutions/2007d\_adv}}, where in addition 
to parameters estimated in the previous solution, we estimated as global 
parameters quantities $\vec{r}_{ok}, \vec{\dot{r}}_{k} $ for all stations, 
quantities $ \vec{h}^c_{ki}, \vec{h}^s_{ki}$ for all VLBA stations and 
25 selected sites, and quantities $ \vec{f}_{kj} $ for five stations.
The polar motion, UT1, and their first time derivatives were also estimated.

\subsection{Required minimum constraints}   \label{s:cns_glo_anal}

  Equations of light propagation are differential equations of the second
order. Their solution does not allow determining specific coordinates of 
sources and stations, but rather a family of coordinate sets. Boundary 
conditions should be formulated either implicitly or explicitly in the form 
of constraints in order to select an element from these sets. These boundary 
conditions cannot in principle be determined from the observations. Thus, 
observations alone are not sufficient to evaluate station positions and 
source coordinates. Coordinates are determined from observations in the 
form of observation equations {\it and\ } boundary conditions in the form 
of constraint equations.

  Expressions for VLBI path delays are invariant with respect to a group of
coordinate transformation that involves translation and rotation of site
positions at a reference epoch, their first time derivatives, and rotation
of source coordinates. In order to remove the rank deficiency, we imposed 
constraints in the form

\begin{eqnarray} 
   \begin{array}{l@{\quad\enskip}l}
     \dss\sum_k^{n_s} (\Delta \vec{r}_{ok} \times \vec{r}_{ok})/|\vec{r}_{ok}| = \const & 
         \dss\sum_k^{n_s}  \Delta \vec{r}_{ok} = \const
     \vex \\
     \dss\sum_k^{n_s} (\Delta \vec{\dot{r}}_{k} \times \vec{r}_{ok})/|\vec{r}_{ok}| = \const & 
         \dss\sum_k^{n_s}  \Delta \vec{\dot{r}}_{k} = \const
     \vex \\
     \dss\sum_a^q \Delta \vec{s}_a \times \vec{s}_a = \const, & \\
   \end{array}
   \label{e:ga2}
\end{eqnarray} 
   where $ \vec{s}_a $ is the coordinate vector of $a$th source, $n_s$ is the 
number of stations that participate in constraints, and $q$ is the number of 
sources that participate in constraints.

  The pairs of parameters $ \vec{r}_{ok}, \vec{f}_{ki}$ and
$ \vec{\dot{r}}_k, \vec{f}_{ki} $ are linearly dependent, and pairs of 
parameters $ \vec{f}_{ki}, \vec{h}^c_{ki}$, and $ \vec{f}_{ki}, \vec{h}^s_{ki}$
may be highly correlated depending on frequencies. In order to avoid rank 
deficiency of a system of observation equations, the following 
decorrelation constraints are to be imposed for each frequency of the 
harmonic constituents: 
\begin{eqnarray}
   \begin{array}{lcr}
     \displaystyle\sum_{j=1-m}^{m-1} f_j \int\limits_{-\infty}^{+\infty} 
                   B_j^m(t) \cos \omega_i \, t \: dt  = \const \vex \\
     \displaystyle\sum_{j=1-m}^{m-1} f_j \int\limits_{-\infty}^{+\infty} 
                   B_j^m(t) \sin \omega_i \, t \: dt  = \const.
   \end{array} 
   \label{e:ga3}
\end{eqnarray}

  Decorrelation constraints between the estimates of B-spline coefficients, 
the estimate of mean site position $\vec{r}_{ok}$ and linear velocity 
$\vec{\dot{r}}_{k}$ (in the case if the degree of B-spline $m>0$) are
to be imposed as well:
\begin{eqnarray}
   \begin{array}{lcr}
     \displaystyle\sum_{j=1-m}^{m-1} f_j \int\limits_{-\infty}^{+\infty} 
                   B_j^m(t) \: dt  = \const \vex \\
     \displaystyle\sum_{j=1-m}^{m-1} f_j \int\limits_{-\infty}^{+\infty} 
                   t \, B_j^m(t) \: dt  = \const.
   \end{array} 
   \label{e:ga4}
\end{eqnarray}

  The integrals \ref{e:ga3}--\ref{e:ga4} can be evaluated analytically 
\citep{r:nur_spl}.

  Similar to coordinates, the adjustments of harmonic variations in coordinates
are invariant with respect to a group of transformations that involve 
translation and rotation. In order to remove the rank deficiency associated
with this group of transformations, we imposed the following constraints:

\begin{eqnarray} 
   \begin{array}{l@{\qquad\quad}l}
     \dss\sum_k^{n_s} ( \vec{h}^c_{ki}  \times \vec{r}_k)/|\vec{r}_k| = \const & 
         \dss\sum_k^{n_h}  \vec{h}^c_{ki}  = \const
         \vex \\
     \dss\sum_k^{n_s} ( \vec{h}^s_{ki}  \times \vec{r}_k)/|\vec{r}_k| = \const & 
         \dss\sum_k^{n_h}  \vec{h}^s_{ki}  = \const.
         \vex \\
   \end{array}
   \label{e:ga5}
\end{eqnarray} 

  In our solutions, we set the constants in equations~\ref{e:ga2}--\ref{e:ga5} 
to zero.

\subsection{Motion of the reference point due to antenna instability} 
\label{s:pisa}

  The residuals of time series of baseline length estimates with station
{\sc pietown} with respect to a linear fit show a significant systematic 
behavior. In our efforts to understand the origin of this behavior, we 
investigated the effect of variations of up to $4'$ in the antenna's tilt,
made evident from pointing measurements.

   When an antenna is pointed at a source, the actual azimuth and
elevation commands sent to the antenna control unit are the expected
azimuth and elevation on the sky, including refraction, plus offsets
that adjust for imperfections in the antenna and encoders.  The
offsets typically amount to a few arc-minutes. They may be calculated using 
pointing equation \ref{e:pointing} in which the imperfections are
parameterized in terms of expected physical effects.  The same
equation, with site specific values for the coefficients, is used at
all VLBA stations.  The coefficients for the different effects in equation
\ref{e:pointing} are determined in pointing observations designed for 
that purpose.  

\begin{eqnarray}
  \begin{array}{r@{\enskip}c@{\enskip}ll}
     \Delta \Az & = & T_e \sin \Az 
                    & + \: T_n \cos \Az 
                    \\ & & 
                      + \: a_0 
                    & + \: a_1 \cos \El 
                    \\ & & 
                      + \: a_2 \sin \El 
                    & + \: a_3 \cos 2\Az 
                    \\ & & 
                      + \: a_4  \sin 2\Az 
                    & + \: H_a(\Az, \El) \vspace{1.0ex} \\
     \Delta \El & = & - \: T_e \cos \Az \, \sin \El 
                    & + \: T_n \sin \Az \, \sin El 
                    \\ & & 
                      + \: e_0 
                    & + \: e_1 \cos \El 
                    \\ & & 
                       + \: e_2 \cos \El 
                    &  + \: e_3 \cos \El \, \sin \Az  
                    \\ & & 
                       + \: e_4 \cos \El \, \cos \Az 
                    &  + \: e_5 \cos \El \, \sin 2\Az 
                    \\ & & 
                       + \: e_6 \cos \El \, \cos 2\Az 
                    &  + \: H_e(\Az ). \\
  \label{e:pointing}
  \end{array}
\end{eqnarray}

  $H_a(\Az, \El)$ and $H_e(\Az )$ are the contribution to azimuth and 
elevation offsets due to rail height variations. The rail height 
H(\Az) was determined for the VLBA antennas by leveling for every 
$3^\circ$ along the circular rail track of 15.24~m diameter. A three-parameter
model $ H_o + H_c \cos \Az + H_s \sin \Az $ was fit to these raw 
measurements and subtracted. The effect of rail height variations is 
complicated by the fact that there are 4 wheels, each responding to the 
rail height at its location and distorting the antenna mount accordingly.   
A simple model, based on analyses by B.~Clark and J.~Thunborg 
(private communication), was developed to describe the effect of rail height:

\begin{eqnarray}
  \begin{array}{ll}
      H_a(\Az, \El) & = \\ 
                    & \hspace{-1em} \hphantom{+}
                        h_{a1} \sin \El \: 
                        \bigl[ H(\Az + 45^\circ) - H(\Az - 45^\circ) \bigr] \\
                    & \hspace{-1em} +
                        h_{a2} \sin \El \: 
                        \bigl[ H(\Az + 135^\circ) - H(\Az - 135^\circ) \bigr] \\
                    & \hspace{-1em} -
                        h_{a3} \cos \El \: 
                        \bigl[ H(\Az + 45^\circ) - H(\Az - 45^\circ) \bigr] \\
                    & \hspace{-1em} +
                        h_{a4} \cos \El \: 
                        \bigl[ H(\Az + 135^\circ) - H(\Az - 135^\circ) \bigr] 
                        \vspace{1ex} \\
      H_e(\Az ) & = \\ 
                & \hphantom{+}
                  h_{e1} \; H(\Az + 135^\circ) \\
                & +
                  h_{e2} \; H(\Az - 135^\circ) \\
                & -
                  h_{e3} \; H(\Az + 45^\circ)  \\
                & -
                  h_{e4} \; H(\Az - 45^\circ).  \\
  \label{e:rails}
  \end{array}
\end{eqnarray}

  It should be noted that the eight coefficients $h_{a1}$---$h_{a4}$ and
$h_{e1}$---$h_{e4}$ are linearly dependent. The equations
\ref{e:rails} can be reduced to linear combinations of two independent 
parameters. Since all VLBA antennas have identical design, these parameters
are considered to be the same for all antennas. They were determined for 
several antennas with the largest rail height variations and then kept fixed.
The method is described in details by \citet{r:cwa_ant}.

  The parameters of pointing equations \ref{e:pointing}, $T_e$, $T_n$, 
$a_0$--$a_4$, $e_0$--$e_6$, are determined using least squares fits 
to measurements of residual pointing offsets. These measurements are made 
at 13 observing bands right after the weekly maintenance day, during 
``startup'' observations designed to help verify proper operation of the 
telescope.  Special targeted pointing observations, often of order 10 hours 
in length, are made during other times that the antennas are not needed for 
interferometer observations.  These special observations often concentrate 
on the 22~GHz and 43~GHz bands. 

  Such measurements are made by recording the total power 
output in baseband channels of 16 MHz bandwidth attached to both left and 
right circular polarization output from the receiver while pointing at each 
of 10 positions near the expected position of a strong source.  The 10
points are off-source, half-power, on-source, half-power, and off-source,
in both azimuth and elevation, with the two half-power and off
positions being on opposite sides of the source.  The off positions
are about 6 beam half-widths from the source, but, for the elevation pattern, 
much of that offset is in azimuth to allow even steps in elevation. The 
full width of half maximum of the beam at 22~GHz is $1'.9$. The residual 
pointing offset and gain are determined by subtracting interpolated 
off-source powers from the on-source and half-power number, and then fitting 
for a peak amplitude and position. The even steps in elevation of the 
elevation scan make the removal of gradients in elevation, such as 
naturally arise from the atmosphere, more effective.

\begin{figure*}[ht]
  \label{f:pietown_pisa}
  \par\smallskip\par
  a) East component  \hspace{0.39\textwidth}
  b) North component
  \par\smallskip\par
  \ifprep
     \includegraphics[width=0.48\textwidth,clip]{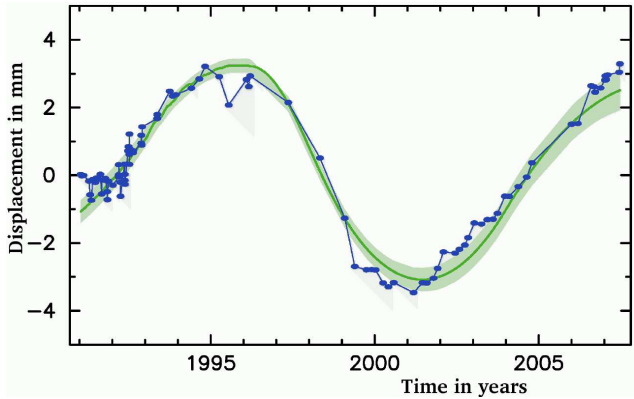}
     \hspace{0.03\textwidth}
     \includegraphics[width=0.48\textwidth,clip]{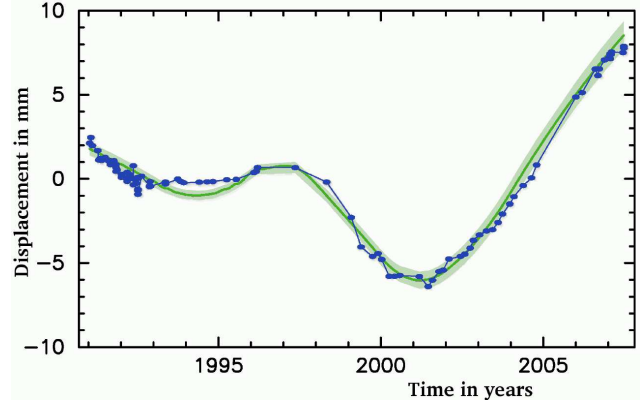}
  \else
     \includegraphics[width=0.48\textwidth,clip]{pietown_tilt_east.eps}
     \hspace{0.03\textwidth}
     \includegraphics[width=0.48\textwidth,clip]{pietown_tilt_north.eps}
  \fi
  \caption{Nonlinear motion of station {\sc pietown} as
           estimated from global VLBI analysis (smooth line) and
           changes in the tilt converted to displacement of the 
           antenna reference point, considering rigid rotation of 
           the antenna. The shadow shows the 1-$\sigma$ formal uncertainties
           of the displacement estimate.}
\end{figure*}

\begin{figure*}[htb]
  \label{f:pietown_gps}
  \par\smallskip\par
  a) East component  \hspace{0.39\textwidth}
  b) North component
  \par\smallskip\par
  \ifprep
      \includegraphics[width=0.48\textwidth,clip]{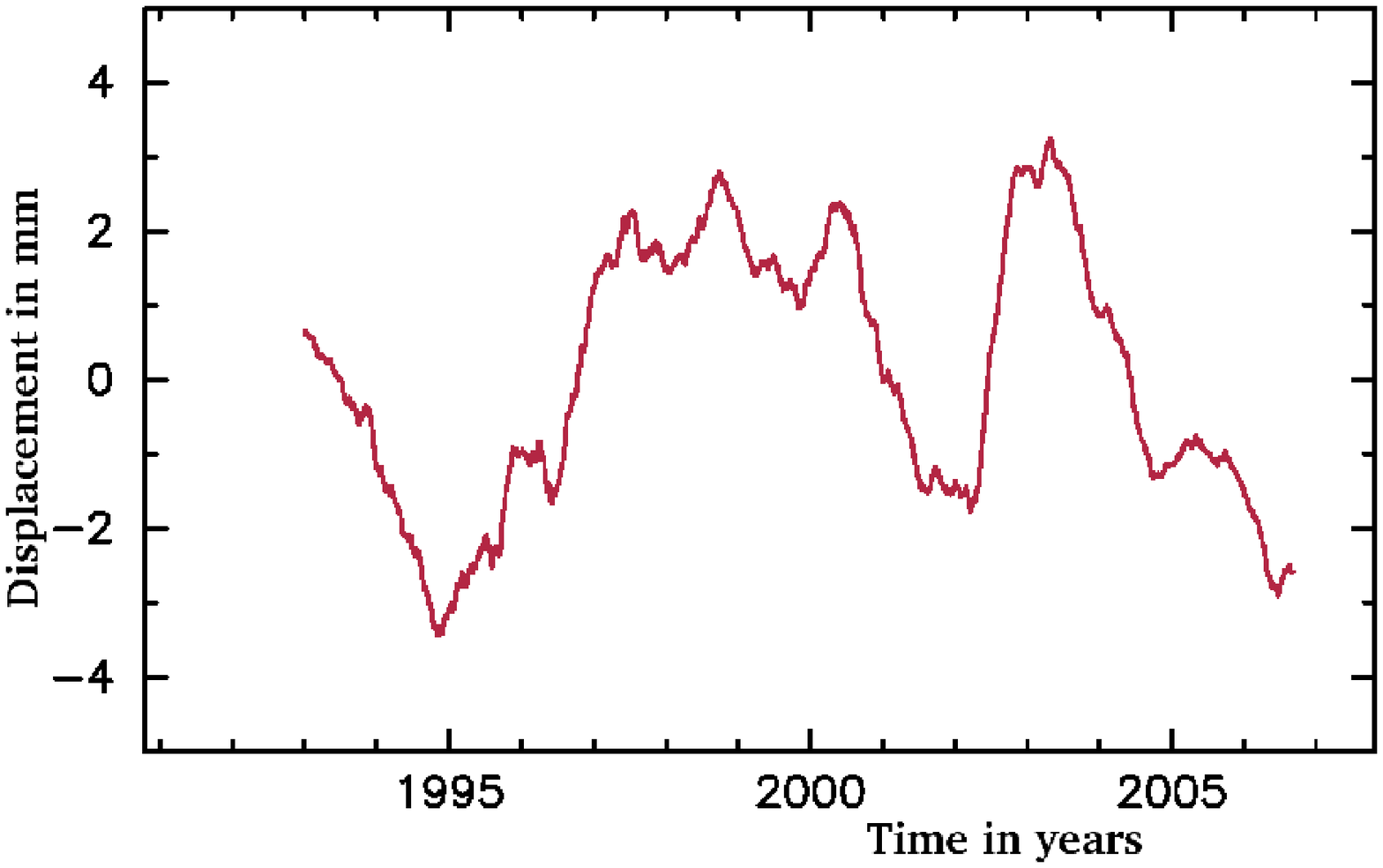}
      \hspace{0.03\textwidth}
      \includegraphics[width=0.48\textwidth,clip]{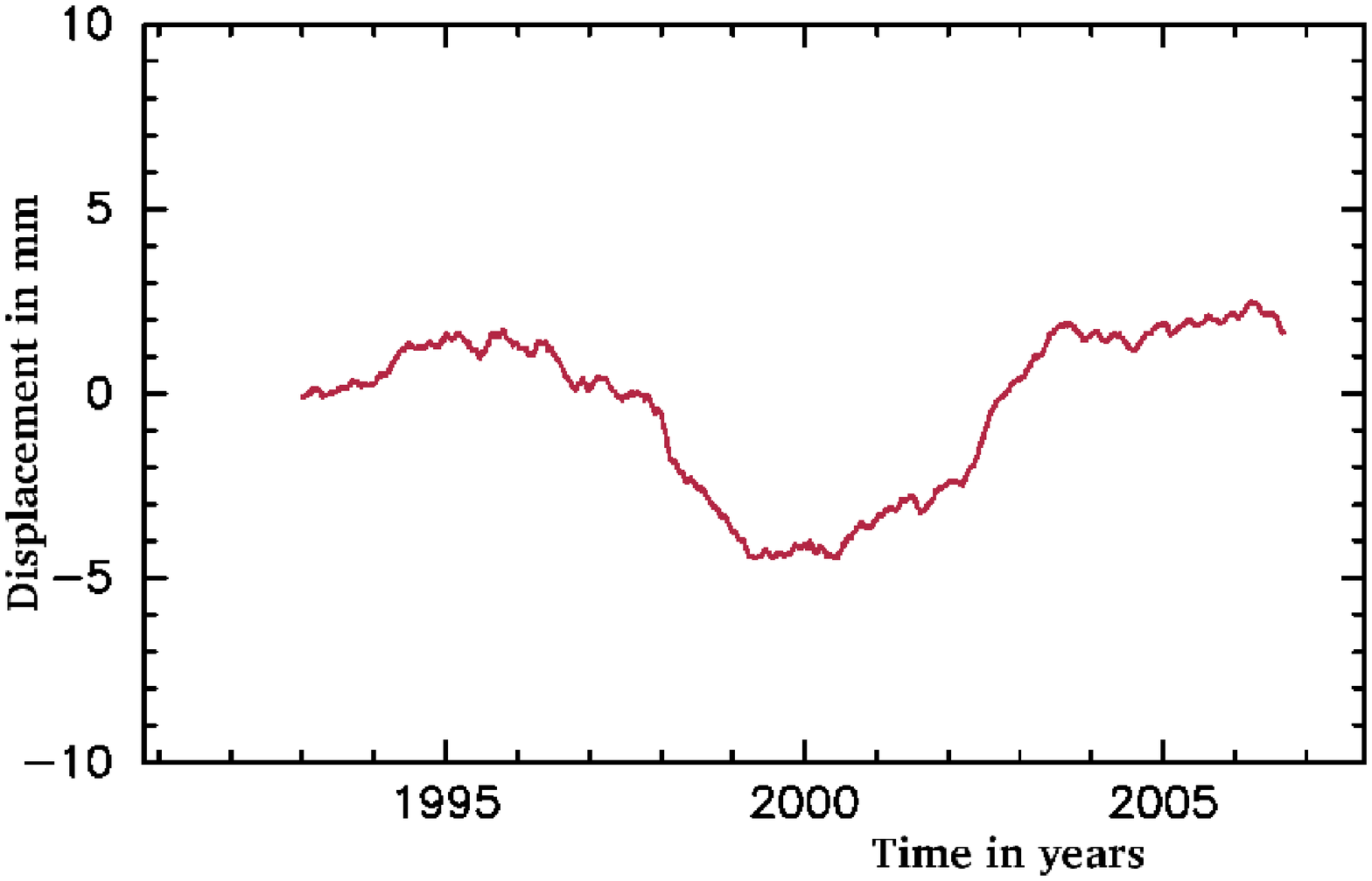}
  \else
      \includegraphics[width=0.48\textwidth,clip]{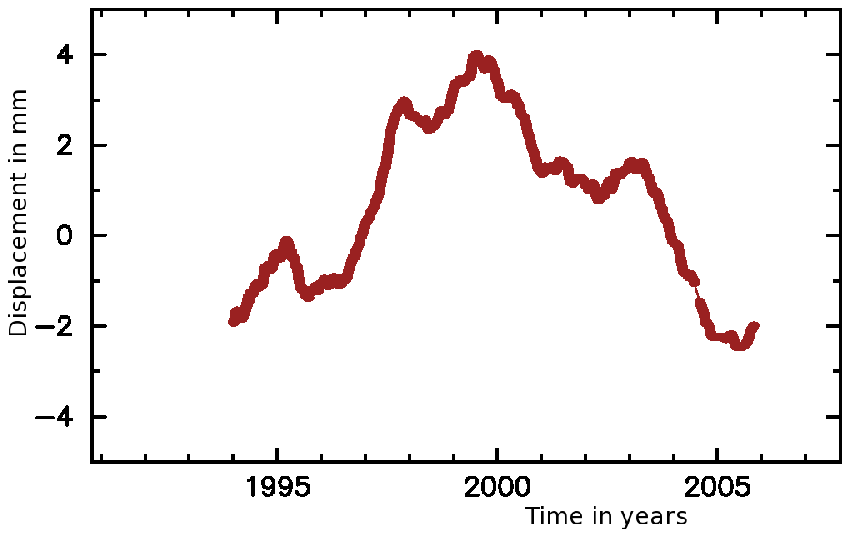}
      \hspace{0.03\textwidth}
      \includegraphics[width=0.48\textwidth,clip]{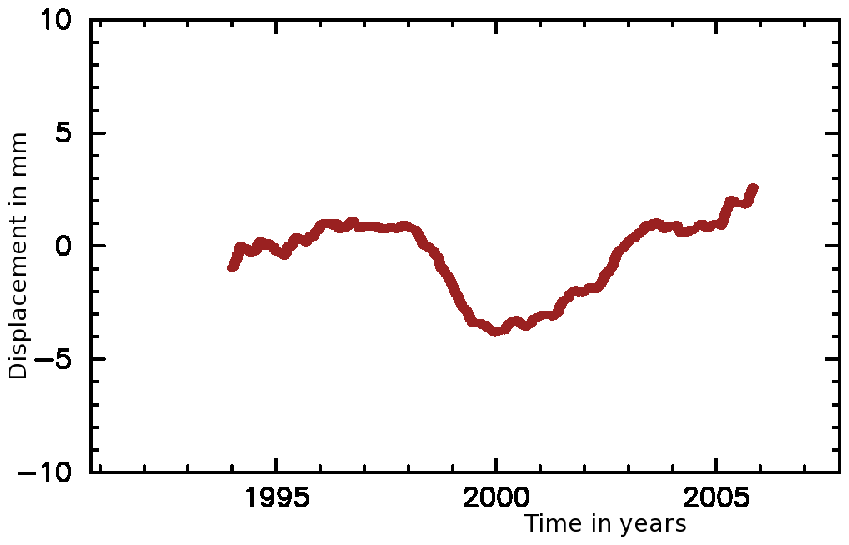}
  \fi
  \caption{Nonlinear motion of station {\sc pie1} as estimated from GPS
           time series after subtracting the mean position, linear velocity,
	   annual and semi-annual position variations from the LSQ
           fit of daily estimates of station positions 
           generated by \citet{r:gps_reanal} and subsequent smoothing.}
\end{figure*}

  Throughout this process, a modulated noise calibration signal of known
equivalent temperature is injected in front of the receiver amplifiers
and synchronously detected along with the total power. This allows
calibration of the total power and the power contributed by the target source 
in terms of antenna temperature. Using sources of known flux density the gain
is determined as a ratio of the source flux density to the antenna temperature. 
Continuum sources used for pointing observations must be stronger than 5~Jy 
to provide enough power so that they are not masked by normal atmospheric 
fluctuations.  

  Occasionally, all pointing observations made over the course of about 
three months are gathered together for a single fit.  Such an analysis 
typically involves about 1000 separate measurements at each of the 22~GHz 
and 43~GHz bands. These high frequency bands are used to determine most 
of the pointing equation parameters because, with their smaller beam widths, 
they produce more accurate pointing measurements.  

  Prior to a fit, the effects of beam squint are removed. This is the offset 
between the left and right circular polarized beams caused by the asymmetric 
geometry of the VLBA antennas. The offset amounts to about 5\% of the beam 
Full Width Half Maximum (FWHM).  An antenna is commanded to point to 
a position half way between the right and left circular polarization beams 
regardless of the polarization being observed, so it is always pointed 
about 2.5\% of the FWHM away from the beam center.

  In the fit for the pointing parameters, some of the terms described
above are held fixed.  The axis non-perpendicularity $a_2$ is generally 
held to zero.  

  Some of the terms of the pointing equation are highly correlated,
so the allocation of offsets to individual physical effects may have
large errors.  But as long as a set of terms determined in a single
solution are used, the derived pointing offsets will be good.  An
example is the constant, sin(El), and cos(El) terms in the elevation
equation (Equation~\ref{e:rails} for $\Delta El$) that have to be 
determined from data that span only about $75^\circ$ in elevation.  
The sum of the effects of these terms is well determined and that is all 
that matters for pointing.  

  The terms  $T_e$ and $T_n$ give the east and north tilts of the fixed 
azimuthal axis. They depend on $\sin(\Az)$ and $\cos(\Az)$. Since the 
measurements cover all azimuths, these estimates are not significantly 
correlated with other terms. A classic example of tilt is the leaning 
tower in Pisa. The tilt can be considered as a small rotation of the 
antenna as a whole with respect to a certain point. This rotation shifts 
the position of the reference point. 

  From our global solution we derived an empirical model of the {\sc pietown} 
reference point position variations by estimating the B-spline coefficients. 
From these coefficients we computed an empirical time series of {\sc pietown} 
displacements. The empirical time series of horizontal displacements were 
fit using time series of the measured north and east tilts derived from 
antenna pointing, and a single admittance factor was adjusted. The estimate 
of the adjusted parameter is $20.5 \pm 0.2$~m. Its physical meaning is the 
distance between the center of rotation that causes the tilt and the 
displacement of the antenna reference point. As Figure~\ref{f:pietown_pisa} 
shows, the empirical model agrees with the tilt measurements within its 
formal uncertainties, at a 0.5~mm level. Thus, the anomalous {\sc pietown} 
horizontal non-linear motion can be explained almost entirely by variations 
in the tilt of the {\sc pietown} antenna. The site position series 
\citep{r:gps_reanal} from the nearby GPS station {\sc PIE1}, located at 
a distance of 61.8~m from the VLBI antenna, shows some similarity 
in non-linear motion in the north component (Figure~\ref{f:pietown_gps}, 
correlation coefficient 0.87), but not in the east component (correlation 
coefficient -0.73). The origin of the non-linear motion of {\sc pietown} 
has not been firmly established, but is thought to be settling of the ground 
beneath the telescope. The antenna is on sloped ground and is leaning into 
the slope.


\subsection{Analysis of the VLBA array velocity field}   \label{s:glo_anal_vel}

  To address the question of stability of the VLBA array, we would like
to determine if any part of the array exhibits only rigid horizontal
motion, i.e. with relative horizontal velocities close to zero.
Since station velocity estimates depend not only on observations but
on constraint equations with an arbitrary right hand side, the estimates of 
motion of the VLBA array as a whole is also subject to an arbitrary
translation and rotation. This means that all velocity vectors of the 
network stations can be transformed as 
\begin{eqnarray}
    \vec{v}_{nk} = \vec{v}_{ok} + \mat{M}_k \, \vec{s},
\label{e:ga6}\end{eqnarray} 
  where $\mat{M}_k$ is the transformation matrix for the $k$th station, 
and $\vec{s}$ is an 6-vector of small arbitrary translation-rotation:
\begin{eqnarray}
    \mat{M}_k & = & \left(
       \begin{array}{r @{\enskip} r @{\enskip}r @{\enskip}r @{\enskip}r @{\enskip} r}
           \hp 1 & \hp 0 & \hp 0 &  0      &  r_{3k} & -r_{2k} \\
           \hp 0 & \hp 1 & \hp 0 & -r_{3k} &  0      &  r_{1k} \\
           \hp 0 & \hp 0 & \hp 1 &  r_{2k} & -r_{1k} &  0      \\
       \end{array}
    \right) \vex \\
    \vec{s} & = & \left( \: T_1      \enskip T_2      \enskip T_3  \enskip
                         \Omega_1 \enskip \Omega_2 \enskip \Omega_3  \:
                         \nonumber \right)^{\top}.
\label{e:ga7}\end{eqnarray} 

  We can find such a vector $\vec{s}$ that the transformed velocity field
will have some desirable properties. This is equivalent to running a new
solution with different right hand sides in constraint 
equations~\ref{e:ga2}--\ref{e:ga5}.

  Equation \ref{e:ga6} is transformed to a local topocentric coordinate system
of the $k$th station by multiplying it by the projection matrix $\mat{P}_k$:

\begin{eqnarray}
    \mat{P}_k \mat{M}_k \, \vec{s} = \vec{u}_k - \mat{P}_k \, \vec{v}_{ok},
\label{e:ga8}\end{eqnarray} 
where $\vec{u}$ is the vector of the station velocity in a topocentric 
coordinate system after the transformation. We can find vector $\vec{s}$ 
from equation~\ref{e:ga8} if we define $\vec{u}_k$ according to the model 
of rigid motion. We split the set of stations into two subsets. The first
set called ``defining'', exhibits only rigid motion, either  horizontal or
vertical, or both. Stations from the second set, called ``free'', have 
non-negligible velocities with respect to the rigid motion.

  We extended our analysis to four non-VLBA stations located in the vicinity 
of VLBA antennas in order to investigate the continuity of the velocity field. 
The set of defining stations was found by an extensive trial. The residual 
velocity of defining stations with respect to the rigid motion was examined. 
We found a set of 7 defining stations for which the residual velocity does 
not exceed $3\sigma$ (Table \ref{t:resvel}). Six stations qualify as 
horizontal defining stations and three as vertical defining stations.

  Using equation \ref{e:ga8} for the horizontal components of the 
6 horizontal defining stations, we set the horizontal components of vector 
$\vec{u}_k$ to zero, and build a system of linear equations. This system
is augmented by adding equations for the vertical components of the 3 
vertical defining stations and we also set the vertical components of vector 
$\vec{u}_k$ to zero. When the number of equations exceeds 6, the system 
becomes redundant. We solve it by LSQ with a full weight matrix $\mat{W}$:

\begin{eqnarray}
   \mat{W} = 
      \left(
           \mat{P}_a \Cov(\vec{v}_o, \trav{v}_o \: ) \, 
           \tram{P}_a \enskip + \mat{A}
      \right)^{-1},
\label{e:ga9}\end{eqnarray} 
where $\mat{P}_a$ is a block-diagonal matrix formed from matrices $\mat{P}_k$,
$\mat{A}$ is a diagonal reweighting matrix with an additive correction to 
weights. Values of \mbox{$(0.4 \: \mbox{mm/year})^2$} for both horizontal and 
vertical velocity components, corresponding to a conservative measure of 
errors, were used in the matrix $\mat{A}$ in our solution.

  Then, transformation \ref{e:ga8} and the rotation to the local topocentric 
coordinate system were applied to both defining and free stations. 
The covariance matrix for velocity estimates of free stations  was computed as:

\begin{eqnarray}
   \begin{array}{l@{\:}c@{\:}l}
      \Cov( \vec{v}_n, \trav{v}_n) & = & \hpl\,
                       \mat{P}_a \Cov(\vec{v}_o,\trav{v}_o) \,
                       \tram{P}_a \enskip \vex \\ & &
                      +\, \mat{P}_a \mat{M} \Cov(\vec{s}, \trav{s} \: ) \,
                          \tram{M} \enskip \tram{P}_a \\
   \end{array}
\label{e:ga10}\end{eqnarray} 
  and for defining stations as
\begin{eqnarray}
   \begin{array}{l@{\:}c@{\:}l}
     \Cov( \vec{v}_n, \trav{v}_n ) & = & \hpl\,
               \mat{P}_a \Cov(\vec{v}_o,\trav{v}_o \: ) \,
                                \tram{P}_a \enskip \vex \\ & &
            +\, \mat{P}_a \mat{M} \Cov(\vec{s},\trav{s} \: ) \,
               \tram{M} \enskip \tram{P}_a 
               \enskip \vex \\ & &
            +\, \mat{P}_a \Cov(\vec{s},\trav{s} \: ) \, 
              \mat{M} \, \mat{W} \Cov(\vec{v}_o,\trav{v}_o) \,
              \tram{P}_a \enskip \vex \\ & &
            +\, \mat{P}_a \Cov( \vec{v}_o,\trav{v}_o) \, \mat{W} \, 
              \tram{M} \, 
              \Cov( \vec{s}, \trav{s} \: ) \, \tram{P}_a .
   \end{array}
\label{e:ga11}\end{eqnarray} 

  The latter expression takes into account statistical dependence of the 
a~priori velocity $\vec{v}_o$ and the vector $\vec{s}$. 

  The results are presented in Table~\ref{t:resvel}. It is remarkable that
there exists a set of 6 stations spread over distances of 1--3 thousand 
kilometers with an average residual horizontal velocity of only 0.2~mm/yr. 

\begin{table}
  \caption{Station local topocentric velocities with respect to the rigid 
           North American plate. Units: mm/yr. The quoted uncertainties are 
           re-scaled 1-$\sigma$ standard errors. The last column indicates 
           whether the station was used as defining for horizontal (h) 
           or vertical (v) motion of the plate.}
  \label{t:resvel}
  \par\vspace{0.5ex}\par

%
%
%
\begin{tabular}{|l| r@{\,}c@{\,}r | r@{\,}c@{\,}r | 
                    r@{\,}c@{\,}r |l|}
   \hline
   Station   & \nnntab{c|}{Up} & \nnntab{c|}{East} & \nnntab{c|}{North} & 
             Def \\
   \hline
   \sc{br-vlba }  &   0.0 &$\pm$&   0.1  &   1.9 &$\pm$&   0.3  &  -0.3 &$\pm$&   0.4  &   v \\
   \sc{fd-vlba }  &   1.2 &$\pm$&   1.0  &   0.4 &$\pm$&   0.2  &  -0.1 &$\pm$&   0.2  &  h  \\
   \sc{hn-vlba }  &   0.2 &$\pm$&   0.3  &  -0.1 &$\pm$&   0.2  &  -0.1 &$\pm$&   0.2  &  hv \\
   \sc{kp-vlba }  &   2.3 &$\pm$&   1.0  &  -0.2 &$\pm$&   0.2  &   0.1 &$\pm$&   0.2  &  h  \\
   \sc{la-vlba }  &   1.3 &$\pm$&   0.8  &   0.0 &$\pm$&   0.3  &   0.0 &$\pm$&   0.2  &  h  \\
   \sc{mk-vlba }  &   0.9 &$\pm$&   1.1  & -55.0 &$\pm$&   1.4  &  52.5 &$\pm$&   1.1  &     \\
   \sc{nl-vlba }  &  -1.1 &$\pm$&   0.5  &   0.0 &$\pm$&   0.2  &   0.1 &$\pm$&   0.3  &  h  \\
   \sc{ov-vlba }  &   2.0 &$\pm$&   0.8  &  -6.0 &$\pm$&   0.3  &   5.0 &$\pm$&   0.4  &     \\
   \sc{pietown }  &   2.1 &$\pm$&   0.9  &  -0.4 &$\pm$&   0.3  &  -1.5 &$\pm$&   0.3  &     \\
   \sc{sc-vlba }  &  -1.2 &$\pm$&   1.2  &  19.0 &$\pm$&   0.8  &   5.3 &$\pm$&   0.4  &     \\
   \sc{gilcreek}  &   3.0 &$\pm$&   1.4  &   4.0 &$\pm$&   0.6  & -10.2 &$\pm$&   0.7  &     \\
   \sc{ggao7108}  &  -0.9 &$\pm$&   0.5  &  -0.4 &$\pm$&   0.3  &  -0.5 &$\pm$&   0.3  &     \\
   \sc{kokee   }  &   2.7 &$\pm$&   1.0  & -54.3 &$\pm$&   1.3  &  52.9 &$\pm$&   1.2  &     \\
   \sc{westford}  &  -0.3 &$\pm$&   0.3  &  -0.1 &$\pm$&   0.2  &   0.0 &$\pm$&   0.2  &  hv \\
   \hline
\end{tabular}

\end{table}

\subsection{Detection of post-seismic deformations}   \label{s:hawaii}

  One possible cause of non-linear site motions is seismic events. These
events are recorded by networks of seismology instruments and their analysis
allows the derivation of additional information, such as timing
of the event, its magnitude, or direction of a slip.
Such information is independent of geodesy
measurements and can be used for verification of our VLBI results.

\begin{figure}
  \label{f:hawaii_eq}
  \par\medskip\par
  \ifprep
     \includegraphics[width=0.48\textwidth,clip]{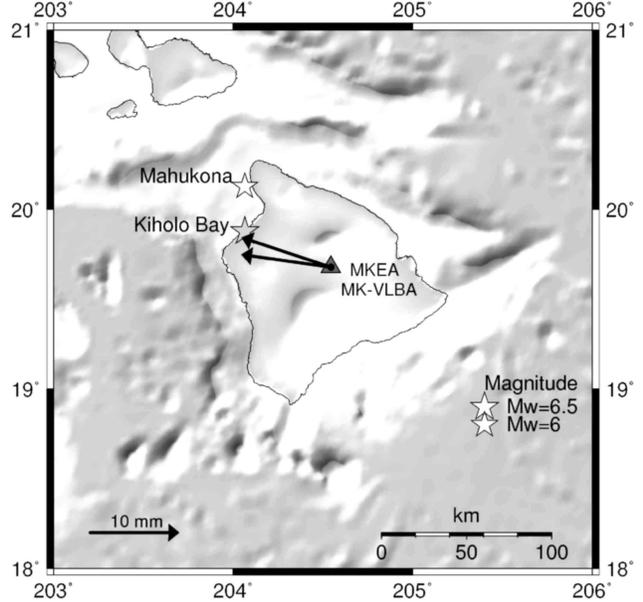}
  \else
     \includegraphics[width=0.48\textwidth,clip]{hawaii_eq.eps}
  \fi
  \caption{Map of the Hawaii Island. Locations of seismic events on 2006.10.15
           are shown with stars. The arrows show the {\sc mk-vlba} site 
           displacement caused by these events from analysis of VLBI 
           observations (bottom) and GPS observations (above).}
  \vspace{-4ex}
\end{figure}

  On 2006.10.15 two powerful earthquakes struck the Island of Hawaii.
A magnitude 6.7 event occurred at 17:07:48 UTC and was located
16~km north-west of Kailua Kona, a town on the west coast of the
Big Island ($19^\circ\!.820$~N, $156^\circ\!.027$~W) in the Kiholo 
Bay, 38 km beneath the surface. The Kiholo Bay event was followed by 
a magnitude \mbox{6.0} Makuhona event 7 minutes later, located 44 km 
north of the airport and at a 20 km depth. The epicenters are shown in 
Figure~\ref{f:hawaii_eq}. Although the two major events
were only 7 minutes apart, their depth difference and aftershock epicenters 
suggest that the second event may not have been an aftershock of the 
larger event, and that they had different sources.  There were no reported 
fatalities, but electric power was lost statewide shortly after the event. 
Despite their moderate depth, the earthquakes generated high accelerations 
in the epicentral region, with strong ground motions lasting for approximately 
20~s during the Kiholo Bay event, and 15~s during the Mahuhona event. 
One station northeast of the epicenter recorded a maximum horizontal
acceleration of 1.03~g. 

  The largest historical earthquakes in Hawaii have occurred beneath the 
flanks of Kilauea, Mauna Loa, and Hualalai volcanoes, when stored compressive 
stresses from magma intrusions into their adjacent rift zones were released.  
Their sources are related to near-horizontal basal decollements at 
approximately 10~km depth, which separate the emplaced volcanic material from 
the older oceanic crust. In contrast, the Kiholo Bay event was considered 
tectonic, rather than volcano related. Deeper, 30--40~km deep, earthquakes 
like this result from a long-term geologic response to flexural fracture of the 
underlying lithosphere from the load of the island mass \citep{r:hawaii}.
They are the result long term accumulation and release of lithospheric flexural 
stresses caused by the island-building process.

  The distance from the epicenter of the Kailua Kona event to the 
{\sc mk-vlba} station is 62 km. No structural damage was reported. 
As a response to the event, an additional observing session was added to the 
schedule on 2006.11.08 in order to detect possible post-seismic deformation. 
Preliminary analysis early in 2007 did not find any position changes 
exceeding 1~cm.

  We reanalyzed the dataset and parameterized non-linear 
motion with the B-spline of the zeroth degree with two knots: 
at 1994.07.08 (beginning observations) and 2006.10.15 17:07:48. Using
these two estimates of B-spline coefficients and their covariance 
matrices, we computed the displacement vector:

\medskip\par
\begin{tabular}{r@{\enskip}c@{\enskip}r@{\,\,}c@{\,\,}l}
   \mbox{Up}    & = & --7.7  & $ \pm $ & 1.3 \,\, \mbox{mm} \\
   \mbox{East}  & = & --10.0 & $ \pm $ & 0.4 \,\, \mbox{mm} \\
   \mbox{North} & = &   1.5  & $ \pm $ & 0.4 \,\, \mbox{mm} 
\end{tabular}

\medskip\par\noindent
  where the quoted uncertainties are unscaled 1-$\sigma$ formal errors.
The vertical and west displacements look very significant. However, the 
presence of outliers may cause an artificial jump and various systematic 
correlated errors may cause estimates of uncertainties to be unreliable. 

  In order to check the robustness of the solution, we performed two tests:
an observation decimation test and a knot shift test. In the observation 
decimation test we ran two solutions. The first solution used only the odd 
observations, while the second solution used only the even observations. 
The data used in these solutions are independent. This test checks the 
contribution of random errors uncorrelated at time scale of the interval 
between observations, typically 5--10 minutes. The difference in estimates 
of the displacement was only 0.14~mm in the vertical component and 0.06~mm 
in the horizontal component. 

  In the knot shift test we made 21 trial solutions which differed only by 
epochs of the B-spline knots. In each trial solution we shifted the epoch of the 
knot six months backward with respect to the previous. The procedure for 
computing theoretical path delay for these trial solutions incorporated 
the estimate of the displacement at 2006.10.15 17:07:48 epoch from the 
initial solution. The rms of the time series of displacement estimates at epochs
with no reported seismic events were 2.1~mm for the vertical and 
1.3~mm for the horizontal component of the {\sc mk-vlba} displacement vector.
We consider these statistics as a measure of the robustness of the estimates 
of the displacement vector. Both tests support our claim that VLBI observations
detected a displacement of station {\sc mk-vlba} caused by the seismic event 
at 2006.10.15 at the confidence level of 99.5\%.

  Since there is a GPS receiver named MKEA located 88 meters from the VLBA 
station {\sc mk-vlba}, we examined the GPS site motion series to determine 
the corresponding co-seismic offset. For the analysis, we used the MKEA daily 
position time series generated by JPL (M.~Heflin, 2007, personal 
communication). We obtained the following estimate for the co-seismic 
displacement vector:
 
\medskip\par
\begin{tabular}{r@{\enskip}c@{\enskip}r@{\,\,}c@{\,\,}l}
   \mbox{Up}    & = & --6.3  & $ \pm $ & 0.9 \,\, \mbox{mm} \\
   \mbox{East}  & = & --9.9  & $ \pm $ & 0.4 \,\, \mbox{mm} \\
   \mbox{North} & = &   3.5  & $ \pm $ & 0.2 \,\, \mbox{mm} 
\end{tabular}
\par\medskip\par

   Here, the uncertainties are unscaled 1-sigma formal errors. This is 
reasonable because the position repeatabilities computed from the 
position time series are close to the formal uncertainties of the daily 
estimates. The VLBI and GPS vertical and east displacements are 
consistent within their 1-sigma error bars but the 2 mm difference in 
the north displacement is too large to be explained by the formal 
uncertainties.

\subsection{Harmonic variations in site positions}   \label{s:harpo}

  The technique of estimation of harmonic variations in site positions 
directly from the analysis of group delays was developed by \citet{r:harpos}.
It was shown in that study that many stations exhibit position variations 
that are attributed to mismodeled harmonic non-tidal signals. The purpose 
of estimating the harmonic site position variations was to remove those
remaining signals. We estimated sine and cosine amplitudes of variations in 
all three components of site position vectors at annual (Sa), 
semi-annual (SSa), diurnal ($S_1$), and semi-diurnal $(S_2)$ frequencies for 
all VLBA and 25 other non-VLBA stations. The seasonal signal is caused by 
unaccounted hydrology loading, by errors in annual amplitudes of the NMF
mapping function that lead to systematic errors in tropospheric path delay
modeling, and possibly other effects. Sun-synchronous diurnal variations 
can be caused by thermal variations, by systematic errors in tropospheric 
path delay, or unmodeled non-tidal ocean loading.

\begin{table}
  \caption{Amplitudes of vertical component of harmonic variations 
           of VLBA station positions in mm.}
  \label{t:uphar}
  \par\vspace{0.5ex}\par
  \begin{tabular}{l @{\qquad}l @{\quad}l @{\quad}l @{\quad}l}
      \hline
       Station          & \nl{annual}  & \nl{semi-annual} & 
                          \nl{\enskip diurnal} & \nl{semi-diurnal} \\ \hline 
       {\sc br-vlba  }  & $  8.0 \pm  0.3 $ & 
                          $  3.3 \pm  0.3 $ & 
                          $  1.0 \pm  0.2 $ & 
                          $  0.5 \pm  0.2 $ \\
       {\sc fd-vlba  }  & $  1.8 \pm  0.3 $ & 
                          $  1.8 \pm  0.3 $ & 
                          $  0.4 \pm  0.2 $ & 
                          $  1.5 \pm  0.2 $ \\
       {\sc hn-vlba  }  & $  5.4 \pm  0.3 $ & 
                          $  2.3 \pm  0.3 $ & 
                          $  1.5 \pm  0.3 $ & 
                          $  1.0 \pm  0.3 $ \\
       {\sc kp-vlba  }  & $  1.7 \pm  0.3 $ & 
                          $  1.8 \pm  0.3 $ & 
                          $  0.2 \pm  0.2 $ & 
                          $  0.8 \pm  0.2 $ \\
       {\sc la-vlba  }  & $  1.0 \pm  0.3 $ & 
                          $  3.2 \pm  0.2 $ & 
                          $  1.3 \pm  0.2 $ & 
                          $  0.9 \pm  0.2 $ \\
       {\sc mk-vlba  }  & $  2.5 \pm  0.4 $ & 
                          $  2.5 \pm  0.4 $ & 
                          $  0.8 \pm  0.3 $ & 
                          $  2.5 \pm  0.3 $ \\
       {\sc nl-vlba  }  & $  4.1 \pm  0.3 $ & 
                          $  3.6 \pm  0.3 $ & 
                          $  0.8 \pm  0.2 $ & 
                          $  0.2 \pm  0.2 $ \\
       {\sc ov-vlba  }  & $  2.7 \pm  0.3 $ & 
                          $  2.0 \pm  0.3 $ & 
                          $  1.1 \pm  0.2 $ & 
                          $  0.9 \pm  0.2 $ \\
       {\sc pietown  }  & $  1.8 \pm  0.3 $ & 
                          $  0.8 \pm  0.3 $ & 
                          $  1.1 \pm  0.2 $ & 
                          $  1.2 \pm  0.2 $ \\
       {\sc sc-vlba  }  & $  2.6 \pm  0.5 $ & 
                          $  3.0 \pm  0.5 $ & 
                          $  0.5 \pm  0.4 $ & 
                          $  1.3 \pm  0.4 $ \\
       \hline
  \end{tabular}
\end{table}

\begin{table}
  \caption{Amplitudes of horizontal component of harmonic variations 
           of VLBA station positions in mm.}
  \label{t:horhar}
  \par\vspace{0.5ex}\par
  \begin{tabular}{l @{\qquad}l @{\quad}l @{\quad}l @{\quad}l}
      \hline
       Station          & \nl{annual}  & \nl{semi-annual} & 
                          \nl{\enskip diurnal} & \nl{semi-diurnal} \\ \hline 
     {\sc br-vlba  }  & $  0.8 \pm  0.1 $ & 
                        $  0.3 \pm  0.1 $ & 
                        $  0.3 \pm  0.1 $ & 
                        $  0.2 \pm  0.1 $ \\
     {\sc fd-vlba  }  & $  1.4 \pm  0.1 $ & 
                        $  0.1 \pm  0.1 $ & 
                        $  0.2 \pm  0.1 $ & 
                        $  0.1 \pm  0.1 $ \\
     {\sc hn-vlba  }  & $  0.9 \pm  0.1 $ & 
                        $  0.2 \pm  0.1 $ & 
                        $  0.2 \pm  0.1 $ & 
                        $  0.1 \pm  0.1 $ \\
     {\sc kp-vlba  }  & $  1.5 \pm  0.1 $ & 
                        $  0.3 \pm  0.1 $ & 
                        $  0.4 \pm  0.1 $ & 
                        $  0.2 \pm  0.1 $ \\
     {\sc la-vlba  }  & $  1.1 \pm  0.1 $ & 
                        $  0.3 \pm  0.1 $ & 
                        $  0.3 \pm  0.1 $ & 
                        $  0.1 \pm  0.1 $ \\
     {\sc mk-vlba  }  & $  0.8 \pm  0.2 $ & 
                        $  0.4 \pm  0.2 $ & 
                        $  0.8 \pm  0.1 $ & 
                        $  0.3 \pm  0.1 $ \\
     {\sc nl-vlba  }  & $  0.7 \pm  0.1 $ & 
                        $  0.2 \pm  0.1 $ & 
                        $  0.2 \pm  0.1 $ & 
                        $  0.2 \pm  0.1 $ \\
     {\sc ov-vlba  }  & $  1.2 \pm  0.1 $ & 
                        $  0.2 \pm  0.1 $ & 
                        $  0.5 \pm  0.1 $ & 
                        $  0.3 \pm  0.1 $ \\
     {\sc pietown  }  & $  1.5 \pm  0.1 $ & 
                        $  0.2 \pm  0.1 $ & 
                        $  0.3 \pm  0.1 $ & 
                        $  0.3 \pm  0.1 $ \\
     {\sc sc-vlba  }  & $  0.7 \pm  0.2 $ & 
                        $  0.9 \pm  0.2 $ & 
                        $  0.4 \pm  0.1 $ & 
                        $  0.3 \pm  0.1 $ \\
       \hline
  \end{tabular}
\end{table}

  In order to evaluate the robustness of the estimates at low frequencies,
we performed two tests: the observation decimation test and the dummy 
frequency test. We examined the differences in estimates of sine and cosine 
amplitudes from the observation decimation test and compared them with the 
formal uncertainties of the estimates. The differences are within 1-sigma 
formal uncertainty. 

  In the second test we estimated site position variations at a frequency of
\mbox{$2.5 \cdot 10^{-7}$} \mbox{rad s${}^{-1}$}, corresponding to a period
of 0.8 year where no harmonic signal is expected. The average amplitude
found for the vertical displacements for the ten VLBA stations is 1.2~mm, and 
0.2~mm for the horizontal displacements. These estimates
should be considered as the upper limit of uncertainties, since the observed 
harmonic signal at frequency \mbox{$2.5 \cdot 10^{-7}$} \mbox{rad s${}^{-1}$} 
is affected by both systematic errors and real displacements at this 
frequency, caused by anharmonic, broad-band site position displacements.

  We see that the combined contribution of seasonal position variation,
unaccounted for in the theoretical model, can reach 1~cm for the vertical 
component of VLBA stations and 1.5~mm for the horizontal component and 
is statistically significant for most of the stations at the 95\% 
confidence level. Unaccounted diurnal position variations are at the 
level of 1--2~mm.

\section{Error analysis} \label{s:errors}

  Uncertainties of estimated parameters can be evaluated using the law of 
error propagation under the assumption that the unmodeled contribution to group 
delay is due to random uncorrelated errors with known variance. The parameter 
estimation procedure provides estimates of these errors based on the SNR of 
fringe amplitudes. These errors are labeled as formal errors and they are 
considered as lower limits of accuracy. Formal uncertainties of the site 
position estimates of the VLBA stations from our global solution are in the 
range of 0.5--1.0~mm for vertical components and 0.2--0.5~mm for horizontal 
components. Formal uncertainties of the VLBA site velocity estimates are in 
the range of 0.07--0.1~mm/yr for vertical components and 0.04--0.05~mm/yr for 
horizontal components. 

  Many factors contribute to an increase of errors. Among them are 
underestimated uncertainties of group delays due to phase instability of 
the data acquisition system, unmodeled instrumental errors, unaccounted 
atmospheric fluctuations, correlations between observations, and unaccounted 
environmental effects.

  Another measure of accuracy is an observation decimation test. Since
the two datasets have independent {\it random} errors, the root mean 
square of differences between estimates from these solutions divided by 
$\sqrt{2}$ provides a measure of accuracy that is independent of
estimates of the uncertainty of each individual observation. 

  However, many other factors that affect the results, such as mismodeled delay 
in the neutral atmosphere, are common in the two subsets. To examine the
influence of these factors, we ran a session decimation test and used every 
second observing session. In the observation decimation test, matrices of 
observation equations were almost identical, but the data were affected by 
the same systematic errors. In the session decimation test, systematic errors 
were more independent, but the matrices of observation equations have larger
differences.

  The statistics of differences are given in Table \ref{t:dec}. In the absence 
of systematic errors, both decimation tests would give close results. Analysis 
of the statistics shows significant discrepancies between the decimation tests. 
Estimates of site positions and velocities in solutions where every second 
observation is removed are a factor of 2--3 closer to each other than in 
solutions where every second session is removed. This is an indication that 
systematic errors on the time scale of several minutes --- the 
typical time between observations --- are correlated. The session decimation
test suggests that estimates of the vertical site position errors should 
be scaled by a factor of 2. This scaling may be related to unaccounted errors 
in modeling the contribution of the neutral atmosphere.

\begin{table}
   \caption{Formal uncertainties and rms of differences of two decimation
            tests for estimates of site positions and site velocities.
            The estimates are given for horizontal and vertical components
            separately.}
   \label{t:dec}
   \begin{tabular}{l@{\quad}ll@{\quad}ll}
      \hline
      Statistics             &  \nntab{c}{Position} & \nntab{c}{Velocity}  \\
                             &  \nntab{c}{mm} & \nntab{c}{mm/yr}           \\
                             &  \ntab{c}{v} & \ntab{c}{h} & 
                                \ntab{c}{v} & \ntab{c}{h}   \\
      \hline
      Formal $\sigma$        &  0.7 & 0.3   & 0.11 & 0.04 \\
      Observation decimation &  0.3 & 0.06  & 0.04 & 0.02 \\
      Session decimation     &  1.4 & 0.3   & 0.09 & 0.06 \\
      \hline
   \end{tabular}
\end{table}

  We also estimated Earth orientation parameters in our solutions. Comparison 
of our EOP estimates with independent GPS time series 
{\tt igs00p03.erp}\footnote{Available at \\ 
ftp://cddisa.gsfc.nasa.gov/gps/products/igs00p03.erp.Z}            
gives us another measure of the accuracy of our results. We computed the rms 
of differences in pole coordinates for sessions in astrometric mode and 
sessions in geodetic modes. Only sessions after 1997 were used for this 
comparison, since GPS estimates prior to this date are not very accurate. 
As we see from Table \ref{t:eop}, the VLBA estimates of pole coordinates from 
geodetic observations are approximately as close to GPS results as ones from 
regular IVS sessions. However, the EOP from astrometric sessions divert from 
the GPS time series by a factor of 2~larger than the EOP from geodetic VLBI 
sessions.

\begin{table}
   \caption{The rms of differences in pole coordinates estimates between
            the VLBI results and the GPS time series {\tt igs00p03.erp}.
            Only data after 1997.0 are used. Comparison is made separately 
            for VLBA sessions in astrometric mode (only 10 VLBA stations) and 
            VLBA sessions in geodetic mode (10 VLBA stations plus 3--10 
            non-VLBA stations).}
   \label{t:eop}
   \begin{tabular}{l@{\quad}ll}
      \hline
      Sessions                      &  X-pole    & Y-pole    \vspace{1ex} \\
      \hline
      VLBA, Astrometric mode        &  0.87 nrad & 1.15 nrad \\
      VLBA, Geodetic mode           &  0.54 nrad & 0.43 nrad \\
      IVS sessions in 2006--2007    &  0.39 nrad & 0.47 nrad \\
      \hline
   \end{tabular}
   \par\vspace{-2ex}\par
\end{table}

\begin{figure}[htb]
  \label{f:basrep}
  \vex
  \includegraphics[width=0.48\textwidth,clip]{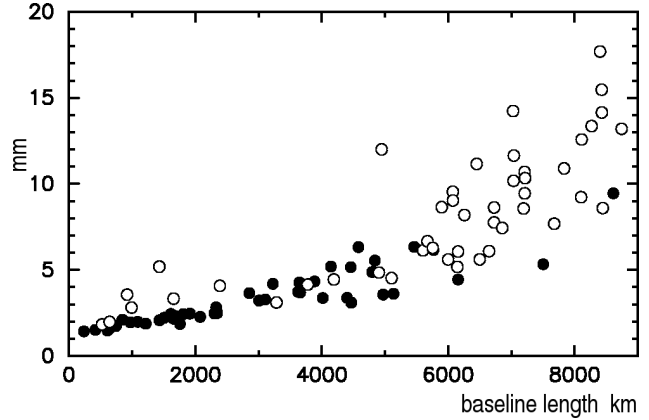}
  \caption{Baseline length repeatability as a function of baseline length.
           Solid disks shows estimates of baseline length repeatability 
           between VLBA sites, circles shows repeatability between 
           non-VLBA sites.}
  \par\vspace{-1ex}\par
\end{figure}

  A baseline length repeatability test provides another measure of solution 
accuracy. For each baseline, a series of lengths was obtained. 
Empirical non-linear site position variations described above were 
applied as a~priori. A plot of the baseline length repeatability of VLBA 
baselines is presented in Figure~\ref{f:basrep}. For comparison, baseline
length repeatability at non-VLBA baselines is also shown.

  A linear model of baseline lengths was fit to each series, and the wrms 
of the deviations from the linear model, the baseline length repeatability,
was computed for each baseline. The plot of baseline length repeatability 
shows that the scatter in baseline lengths estimates between VLBA sites 
is less than the scatter in baseline length between dedicated geodetic 
VLBI stations. The set of wrms was fit by a function 
$ \sqrt{ A^2 + (B \cdot L)^2} $ where $L$ is the mean baseline length. 
Coefficients $A$ and $B$, which represent the average baseline length 
repeatability, are a measure of accuracy. For the VLBA baselines,
A = 1.6~mm, B=0.9~ppb, for non-VLBA baselines A=2.0~mm, B=1.4~ppb. Growth
of the baseline length repeatability with the baseline length for both 
sets of data reflects the impact of the contribution of unmodeled path delay 
in the neutral atmosphere, which affects the site position vertical component 
to a greater extent than the horizontal one \citep{r:davis}.

\begin{figure}[htb]
  \label{f:pos_error}
  \vex
  \includegraphics[width=0.48\textwidth,clip]{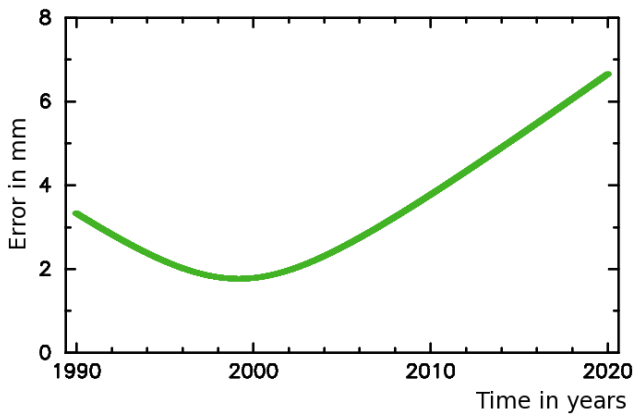}
  \caption{Predicted uncertainties of {\sc mk-vlba} vertical coordinate if
           no additional data are taken after 2007. The errors were inflated 
           by a factor of 2 with respect to the formal errors. 
           The uncertianties of positions of other stations 
           have a similar growth.}
  \par\vspace{-2ex}\par
\end{figure}

  The results of error analysis allow us to conclude that the 
errors of predicted site positions for any epoch within the time range of 
observations, [1994, 2008], are in the range of 2--3~mm for the vertical 
component and 0.4--0.6~mm for the horizontal component. The predicted 
positions, based on the adjusted parameters of the site motion model,
includes mean site positions at the reference epoch, site velocities, and
coefficients of the harmonic and B-spline models. The estimates of errors 
for the vertical coordinates were derived from the formal errors by inflating 
by a factor of 2, as the session decimation test suggests. In the absence of
new observations, the predicted errors in site position will grow by a factor 
of 2 by 2020 as is shown in Figure~\ref{f:pos_error}, provided no motion 
other than harmonic position variations and linear velocities, i.e. that
no seismic events, or unmodeled variable tilt, will happen in the future.

\section{Conclusions} \label{s:conclusions}

  The observing campaign for monitoring positions of the VLBA sites 
during 1994--2007 has been highly successful. From analysis of 14~years 
of data, the elements of the VLBA array during that period were determined 
with an accuracy of 2--3~mm in the vertical and 0.4--0.6~mm in the 
horizontal component. This meets the requirements for position accuracy 
for astrometry and astrophysics programs at the VLBA.

  The baseline length repeatability between VLBA stations is smaller than
that for non-VLBA IVS stations. EOP estimates from geodetic VLBA 
sessions are as close to GPS results as EOP estimates from the IVS sessions 
dedicated to precise EOP determination. 

  We found that the positions of all VLBA stations exhibit a significant 
seasonal signal with amplitudes of 1--8~mm in the vertical and
0.5--3.5~mm in the horizontal component. 

  Several stations show anharmonic signals in their positions. We have traced
the origin of these signals to co-seismic deformations ({\sc mk-vlba}) and 
to a time-varying antenna tilt ({\sc pietown}). In the case of tilt, the 
signal can be successfully modeled using the pointing adjustment model, 
however, the scaling factor between the antenna tilt and motion of the 
antenna reference point has to be determined from VLBI observations.

  We derived an empirical model of site motion that consists of linear
velocity, a set of coefficients of the harmonic expansion, and coefficients 
of the B-spline model that takes into account ad~hoc motions. For the case 
where no ad~hoc motion occurs in the future, the accuracy of VLBA station 
position predictions would gradually degrade to 5--8~mm in the vertical 
and 1--1.5~mm in the horizontal by the year 2020 in the absence
of future observations. However, unpredictable events, such as local 
deformations or post-seismic deformation could cause significantly larger 
errors. Therefore, continuation of VLBA site position monitoring is highly 
desirable.

\begin{acknowledgements}

  The National Radio Astronomy Observatory is a facility of the National 
Science Foundation operated under cooperative agreement by Associated 
Universities, Inc. We thank the staff of the VLBA for carrying out and
correlating these observations in their usual efficient manner. We also thank
M.~Titus and B.~Corey from Haystack Observatory for their efforts in 
re-correlation of the rdv22 data on the Haystack correlator. Lastly, we are 
thankful to A.~Nothnagel and A.~Niell for valuable comments that helped 
to improve this manuscript. This work was done while J.~Gipson, D.~Gordon, 
D.~MacMillan and L.~Petrov worked for NVI, Inc. under NASA contract 
NAS5--01127.

\end{acknowledgements}

\par\vspace{-2ex}\par

\end{document}